\documentclass{pazhb}
\usepackage[hyperfootnotes=false]{hyperref}
\usepackage{graphicx}
\usepackage{color}
\usepackage{epsf}
\usepackage[utf8]{inputenc}
\usepackage[T2A]{fontenc}
\usepackage{amsmath}
\usepackage{amsfonts}
\usepackage{amssymb}
\usepackage{epsf}
\usepackage{gensymb}
\usepackage{float}
\usepackage{longtable}
\usepackage{changes}
\usepackage{wrapfig}
\usepackage{xcolor}

\newcommand{\msun}{\mbox{M$_{\odot}$}}

\newcommand{\h}{$^\mathrm{h}$}
\newcommand{\m}{$^\mathrm{m}$}

\newcommand{\fgl}{4FGL~J1943.9+2841}

\newcommand{\sw}{\textit{Swift}}
\newcommand{\xmm}{\textit{XMM-Newton}}
\newcommand{\rosat}{\textit{ROSAT}}
\newcommand{\eros}{eROSITA}
\newcommand{\gaia}{\textit{Gaia}}
\newcommand{\degs}{\ifmmode ^{\circ}\else$^{\circ}$\fi}
\newcommand{\amin}{\ifmmode ^{\prime}\else$^{\prime}$\fi}
\def\asec{\ifmmode ^{\prime\prime}\else$^{\prime\prime}$\fi}

\newcommand{\nh}{$N_\mathrm{H}$}
\newcommand{\ergs}{erg~s$^{-1}$}
\newcommand{\flux}{erg~s$^{-1}$~cm$^{-2}$}
\newcommand{\masyr}{mas~yr$^{-1}$}

\newcommand{\magup}{^{\mathrm{m}}}

\begin{document}

\journalinfo{2024}{50}{6}{351}[372]


\title{SRGe J194401.8+284452 —-- an X-ray Cataclysmic Variable in the Field of the Gamma-Ray Source 4FGL J1943.9+2841
}
\author{
A. I.~Kolbin\address{1,2,9}\email{kolbinalexander@mail.ru},
A.V.~Karpova\address{4},
M.V.~Suslikov\address{1,2},
I.F.~Bikmaev\address{2,3},\\
M.R.~Gilfanov\address{5,6},
I.M.~Khamitov\address{2}, 
Yu.A.~Shibanov\address{4}, 
D.A.~Zyuzin\address{4},  
G.M.~Beskin\address{1,2},\\
V.L.~Plokhotnichenko\address{1},
A.G.~Gutaev\address{1,2},
S.V.~Karpov\address{8},
N.V.~Lyapsina\address{1},
P.S.~Medvedev\address{5},
R.A.~Sunyaev\address{5,6},
A.Yu.~Kirichenko\address{7,4},
M.A.~Gorbachev\address{2},
E.N.~Irtuganov\address{2},
R.I.~Gumerov\address{2},
N.A.~Sakhibullin\address{2,3},
E.S.~Shablovinskaya\address{1},
E.A.~Malygin\address{1}
\addresstext{1}{Special Astrophysical Observatory of RAS, Nizhnij Arkhyz,  Karachai-Cherkessian Republic, 369167, Russia}
\addresstext{2}{Kazan (Volga-region) Federal University, Kremlevskaya 18, Kazan 420008, Russia}
\addresstext{3}{Tatarstan Academy of Sciences, Bauman 20, Kazan 420111, Russia}
\addresstext{4}{Ioffe Institute, Politekhnicheskaya 26, St. Petersburg 194021, Russia}
\addresstext{5}{Space Research Institute, Russian Academy of Sciences, Profsoyuznaya 84/32, 117997 Moscow, Russia}
\addresstext{6}{Max Planck Institute for Astrophysics, Karl-Schwarzschild-Strasse 1, Garching bei München D-85741, Germany}
\addresstext{7}{Instituto de Astronomía, Universidad Nacional Autónoma de México, Apdo. Postal 877, Ensenada, Baja California 22800, México}
\addresstext{8}{CEICO, Institute of Physics, Czech Academy of Sciences, Prague 18221, Czech Republic}
\addresstext{9}{North Caucasian Federal University, Pushkina 1, Stavropol 355017, Russia}}

\shortauthor{Kolbin et al.}
\shorttitle{Intermediate polar SRGe J194401.8+284452}

\begin{abstract} 

SRGe J194401.8+284452 is the brightest point-like X-ray object within the position uncertainty ellipse of an unidentified $\gamma$-ray source 4FGL J1943.9+2841. We performed multi-wavelength spectral and photometric studies to determine its nature and possible association with the $\gamma$-ray source. We firmly established its optical counterpart with the Gaia based distance of about 415 pc. Our data show that the object is a cataclysmic variable with an orbital period of about 1.5 hours. SRGe J194401.8+284452 exhibits fast spontaneous transitions between the high and low luminosity states simultaneously in the optical and X-rays, remaining relatively stable between the transitions on scales of several months/years. This can be caused by an order of magnitude changes in the accretion rate. The brightness of the source is about 17 and 20 mag in optical range and about $5\times 10^{-12}$ and $5\times 10^{-13}$ erg/cm$^2$/s in the 0.3 -- 10 keV range in the high and low states, respectively. We constrained the mass of the white dwarf (0.3 -- 0.9 $M_\odot$) and its temperature in the low state (14750 $\pm$ 1250 K), the mass of the donor star ($\leq$ 0.08 $\pm$ 0.01 $M_ \odot$). In the low state, we detected regular optical pulsations with an amplitude of 0.2 mag and a period of 8 min. They are likely associated with the spin of the white dwarf, rather than with its non-radial pulsations. In the high state, the object demonstrates only stochastic optical brightness variations on time scales of 1 -- 15 minutes with amplitudes of 0.2 -- 0.6 mag. We conclude, that SRGe J194401.8+284452 can be classified as an intermediate polar, and its association with the $\gamma$-ray source is very unlikely.

an accretion disk around the white dwarf. 

\end{abstract}

\section{Introduction}

The X-ray source XMMSL2 J194402.0+284451 (hereinafter J1944) from the \xmm\ catalog \citep{xmmsl} has the observed flux $f_X=(1.25\pm0.45)\times10^{-11}$~\flux\ in the 0.2 -- 12 keV range according to data from May 3, 2013.
It was previously discovered in the \rosat\ all-sky survey \citep{rosat}, in which this source is designated as 2RXS J194401.4+284456 (53 counts were obtained in 387 sec in October 1990). The nature of J1944 has not yet been established. 

 \citet{torres} suggested that the source is a pulsar wind nebula (PWN) candidate.
Its optical counterpart was also identified by positional coincidence 
with a star from the \gaia\ catalog with a brightness of $G=17.6$\m. According to the results of X-ray and ultraviolet (UV) observations with the \sw\ telescope in 2018 (with exposure time of 850 sec), the source was too faint to be a PWN, and the question about its nature remained open.

In the subsequent work by \citet{Takata22} J1944 was considered as a possible cataclysmic variable (CV). 
The authors used optical data from the TESS telescope, but they were unable to determine the parameters of the proposed binary to confirm this hypothesis.

\begin{figure}
\begin{minipage}[h]{1.\linewidth}
\center{\includegraphics[width=1.0\linewidth,clip]{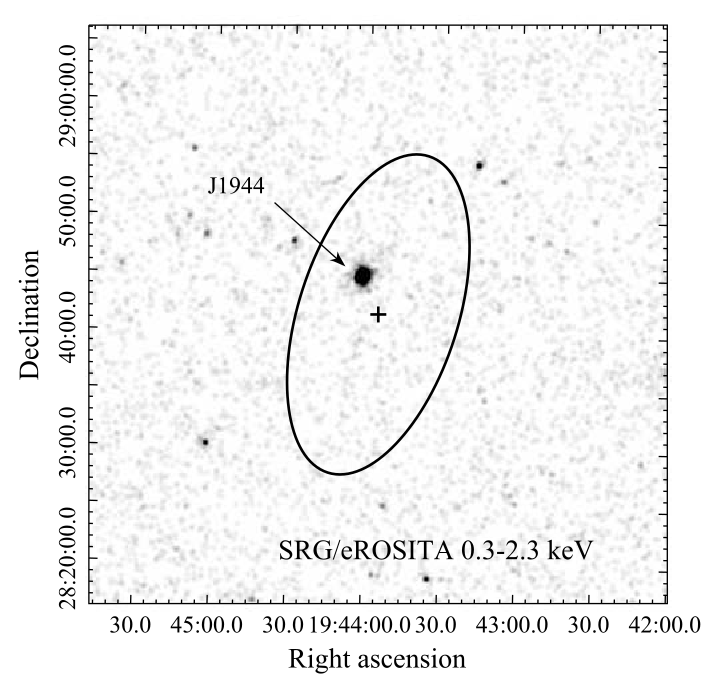}}
\end{minipage}
\caption{50\amin $\times$ 50\amin\ \eros\ image of the J1944 field 
in the 0.3 -- 2.3 keV range.
The position of the $\gamma$-ray source \fgl\ is marked with a cross and the ellipse shows its 95\% uncertainty\citep{4fgl-dr3}.
}
\label{fig:eROSITA}
\end{figure}

Alternatively, J1944 might  be a counterpart  of an unassociated $\gamma$-ray source \fgl. The detection significance of the latter is only $\approx$6$\sigma$ and no brightness variability was detected.
Its $\gamma$-ray flux is $F_\gamma$ = $(4.3 \pm 1.1)\times 10^{-12}$ \flux~\ in the 0.1 -- 100 GeV range and its spectrum can be described by the LogParabola model \citep{4fgl-dr3,4fgl-dr4}. J1944 is the brightest X-ray source in the ellipse, corresponding to the 95\% position uncertainty of \fgl\ (Fig.~\ref{fig:eROSITA}). If the identification of J1944 with the $\gamma$-ray source is confirmed, it could be a pulsar in a binary system.

To clarify the nature of J1944, we examined X-ray data from the eROSITA telescope \citep{predehl2020} on board the Spektr-RG observatory \citep[SRG;][]{sunyaev2021} obtained during the all-sky scanning in 2020--2021 (the corresponding source name is SRGe J194401.8+284452), optical data from various catalogs and archival UV and X-ray data from \sw\ observations (name in the \sw-XRT Point Source catalog -- 2SXPS J194401.7+284450, \citealt{2sxps}).
Optical spectroscopic and photometric observations were also carried out with the 6-m BTA telescope, the 1.5-m Russian-Turkish telescope RTT-150 and the 2.1-m telescope of the Mexican National Astronomical Observatory (Observatorio Astron\'omico Nacional San Pedro M\'artir, OAN-SPM).
Here we present the results of the analysis of these data.

\section{Observations and data reduction}


\subsection{Optical data from catalogs}

The parameters of the J1944 optical counterpart candidate, taken from the \gaia\ catalog \citep{gaia2016,gaia2023}, are presented in Table~\ref{tab:pars}. We found the source in the Isaac Newton Telescope Photometric H-Alpha Survey \citep[IPHAS;][]{iphas}, where 
its apparent magnitudes are $r$ = 19.60(2)\m, $i$ = 19.54(6)\m\ and $H\alpha$ = 18.41(2)\m\ in the AB photometric system on MJD 53563.

There are also IPHAS observations on MJD 54284 and 54674, but their results were not included in the survey, possibly due to non-photometric weather conditions. However, analyzing the corresponding images we found
that on these dates the source had approximately the same brightness as in observations on MJD 53563. 
The best signal-to-noise ratio was obtained in observations 
on MJD 54674 and we calculated its magnitudes of $r$ = 19.90(4)\m\ and $H\alpha$ = 18.66(4)\m.

In addition, we used data from the Panoramic Survey Telescope and Rapid Response System \citep[Pan-STARRS;][]{ps} and Zwicky Transient Facility catalogs \citep[ZTF;][]{ztf} which show significant brightness variability of the source.

\begin{table}
\renewcommand{\arraystretch}{1.2}
\caption{Parameters of the optical counterpart candidate of J1944 from the \gaia\ catalog.
The distance $D$ corresponds to the geometric distance according to \gaia\ \citep{BailerJones}.}
\label{tab:pars}
\begin{center}
\begin{tabular}{lc}
\hline
R.A. (J2000)                                        & 19\h44\m01\fs898273(4)   \\
Dec. (J2000)                                        & +28\degs44\amin51\farcs58412(6) \\
Gal. longitude $l$, deg.                            & 64.226 \\
Gal. latitude $b$, deg.                             & 2.396 \\
\hline
Parallax $\pi$, mas                                  & 2.405(75) \\	
Proper                                             & 35.766 \\	
motion $\mu$, \masyr\                               &     \\	
$\mu_{ra}$, \masyr\                                   & 32.56(6)	\\
$\mu_{dec}$, \masyr\                                  & 14.81(7) \\
Distance $D$, pc                                      & 415(15) \\
\hline
\end{tabular}
\end{center}
\end{table}

\subsection{Optical photometry}

Photometric observations of the J1944 optical counterpart candidate were carried out in the $g$, $r$, $i$ and $z$ filters on 2022, 2023 and 2024 with the 1.5-m Russian-Turkish telescope RTT-150 (T\"UB\.ITAK National Observatory, Antalya, Turkey). The telescope was equipped with a TFOSC instrument and an Andor iKon-L 936 BEX2-DD-9ZQ CCD with a size of $2048\times 2048$ pixels, thermoelectrically cooled to $-80$ C$^{\circ}$. The field of view in the direct image mode is 11\amin$\times$11\amin\ with the pixel scale 0.33\asec\ and binning 1$\times$1.

The object was also observed with the 2.1-m OAN-SPM telescope using  the ``Rueda Italiana'' instrument\footnote{\url{https://www.astrossp.unam.mx/en/users/instruments/ccd-imaging/filter-wheel-italiana}} in the $V$ band on June 1, 2023. The field of view of the detector is 6\amin$\times $6\amin\ with a pixel scale of 0.34\asec\ and binning 2$\times$2.

Photometric data were reduced using the IRAF package\footnote{The IRAF astronomical data processing and analysis package is available at \url{https://iraf-community.github.io}}. The direct images were bias subtracted and flat-field corrected. Cosmic rays were removed using the LaCosmic \citep{Dokkum} algorithm. Source detection in images was performed using the DAOFIND algorithm. Since wings from the nearest star are superimposed on the profile of the J1944 candidate, the PSF photometry was used to correctly calculate the fluxes.


We also carried out fast photometry with the 6-m BTA telescope of the Special Astrophysical Observatory of the Russian Academy of Sciences in 2022. These observations used the MANIA complex based on a multimode panoramic photopolarimeter (MPPP) with microsecond time resolution \citep{Plohotnichenko2021,Plohotnichenko2023}. The instrument was installed at the Nasmyth1 BTA focus, equipped with a rotary table that compensated the field of view rotation with a diameter of about 20\asec. The emission of this sky region in the range 3600 -- 7600 \AA, divided at a wavelength of 4500 \AA\ into two beams by a dichroic mirror, was synchronously recorded by coordinate-sensitive detectors (CSD) with a sensitivity maximum in the red and blue parts of the spectrum \citep{Debur2003,Plohotnichenko2009}. The coordinates and arrival times of photons were detected and encoded by the multidimensional chronometric device ``Kvantokhron 4-48'' \citep{Plohotnichenko2009} with a time resolution of 1 $\mu$s and stored as a data array in the computer memory. The data arrays were converted into photon lists containing sequences of arrival times and coordinates of the detected photons. Using the photon lists, the dynamic images of the MPPP field of view and the light curves of selected sources (the object, a reference star, and two check stars) were constructed, and the statistical parameters were determined.
At the same time, the effects of the atmospheric instability, variations in the telescope guiding speed, as well as the uneven movement of the rotary table were compensated.


The log of all observations is presented in Table~\ref{table:log}.


\begin{table*}
\caption{Observation log of the J1944 optical counterpart candidate. The telescopes and instruments used in the observations, date and duration, number of images obtained (N), exposure time, and spectral range are presented.}
\begin{center}
\label{table:log}
\begin{tabular}{cccccc}
\hline
Telescope/Instrument   & Date   & Duration & N & Exposure & Spectral range \\
                    & (UT)              & HJD$-$2459700       &  & sec &         \\ \hline
RTT-150/TFOSC       & 13/14-05-2022       &  13.4848--13.5940 & 62 & 120 & $g$, $r$, $i$         \\
RTT-150/TFOSC       & 15/16-05-2022       & 15.4831--15.5865 & 44 & 180 & $r$          \\
RTT-150/TFOSC       & 01/02-06-2022       & 32.4337--32.5634 & 63 & 180 & $i$ \\
BTA/SCORPIO-2       & 04/05-07-2022      & 65.3705--65.4836  & 28 & 300 & 3650--7250\AA \\
BTA/SCORPIO-2       & 07/08-07-2022      & 68.4214--68.4892  & 17 & 300 & 3650--7250\AA \\
BTA/MANIA           & 06/07-07-2022      & 67.3822--67.4100  & 3230 & 1 & 3600--7600\AA \\
BTA/MANIA           & 06/07-07-2022      & 67.4197--67.5000  & 7999 & 1 & 3600--7600\AA \\
\hline
2.1-m/Rueda Italiana & 01/02-06-2023       & 396.8990--396.9669 & 51 & 100 & $V$ \\
RTT-150/TFOSC       & 15/16-06-2023       & 411.4484--411.5523 & 20 & 120 & $r$          \\
RTT-150/TFOSC       & 11/12-07-2023       & 437.4336--437.5480 & 90 & 60 & $g$          \\    
RTT-150/TFOSC       & 12/13-07-2023       & 438.4625--438.5642 & 76 & 120 & $r$          \\
RTT-150/TFOSC       & 15/16-07-2023       & 441.4186--441.5569 & 80 & 120 & $r$          \\
\hline
RTT-150/TFOSC       & 07/08-06-2024       & 769.4924--769.5603 & 70 & 60 & $g$          \\
RTT-150/TFOSC       & 09/10-06-2024       & 771.4218--771.5535 & 140 & 60 & $g$          \\
RTT-150/TFOSC       & 15/16-06-2024       &777.4203--777.4512 & 12 & 180 & $g$, $r$, $i$, $z$
\\
\hline
\end{tabular}
\end{center}
\end{table*}



\subsection{Optical spectroscopy}

Spectral data of the 
presumed optical counterpart of J1944 were obtained with the 6-m BTA telescope of the Special Astrophysical Observatory of the Russian Academy of Sciences on July 4/5 and July 7/8, 2022 in the director’s reserve quota. The observations were carried out under good astroclimatic conditions with a seeing of $1.2-1.3$\asec. The spectra were obtained using the SCORPIO-2 focal reducer in the long-slit spectroscopy mode \citep{Afan}. 
The VPHG1200@540 volume phase holographic grating with a 1\asec\ slit width were used, providing the spectral resolution $\delta \lambda \approx 5.5$~\AA\ in the  
$\lambda\lambda = 3650 - 7250$~\AA\ range. 
The spectra were reduced using the IRAF package. Spectral images were bias subtracted, cleared of cosmic rays using the LaCosmic program, and corrected for inhomogeneous sensitivity by flat-field division. Using He-Ne-Ar lamp images, the spectra were wavelength calibrated and corrected for geometric distortions. Optimal extraction of spectra with subtraction of the sky background \citep{Horne86} was performed. The source spectra were flux calibrated using the spectra of the standard star BD+25$^{\circ}$4655 \citep{Oke90}. We also calculated heliocentric radial velocity corrections for each spectrum. The log of spectral observations is included in Table \ref{table:log}. 


\subsection{Archival UV observations}

The J1944 field was observed with the \sw\ telescope in 2018 and 2022 -- 2024. We analyzed Ultra-violet/Optical Telescope (UVOT) data obtained in the $uvw2$ (1928\AA), $uvm2$ (2246\AA), $uvw1$ (2600\AA) and $u$ (3465\AA) bands using processed images from the archive.
For photometry we used the UVOTSOURCE command and a standard aperture of 5\asec. The list of observations and photometry results are  presented in Table~\ref{tab:swift-uvot}. 
The J1944 flux in the $u$ band is contaminated by the light from nearby stars (Fig.~\ref{fig:opt-uv}), while this is negligible at shorter wavelengths. 
Observations 03111808002 and 00010695019 were excluded from consideration because they had UVOT exposures of 7 and 0 sec, respectively.


\subsection{X-ray data}

The J1944 field was observed with the \eros\ telescope with the total exposure time of 669 sec. To extract the source spectrum, we used a circle with a radius of 60\asec, 
and the background spectrum was extracted from an annulus with radii of 120\asec\ and 300\asec.

J1944 was also monitored with the X-Ray Telescope (XRT) 
on board the \sw\ observatory. Observations analyzed in this work are listed in Table~\ref{tab:swift-uvot} with addition of ObsIDs 03111808002 and 00010695019.
\sw-XRT data products generator tools\footnote{\url{https://www.swift.ac.uk/user\_objects/}} \citep{evans2009}, were used to obtain the light curve and to extract spectra of the source.


\section{Data analysis and results}


\subsection{Optical and UV images}

\begin{figure*}
\begin{minipage}[h]{1.\linewidth}
\center{\includegraphics[width=0.48\linewidth,clip]{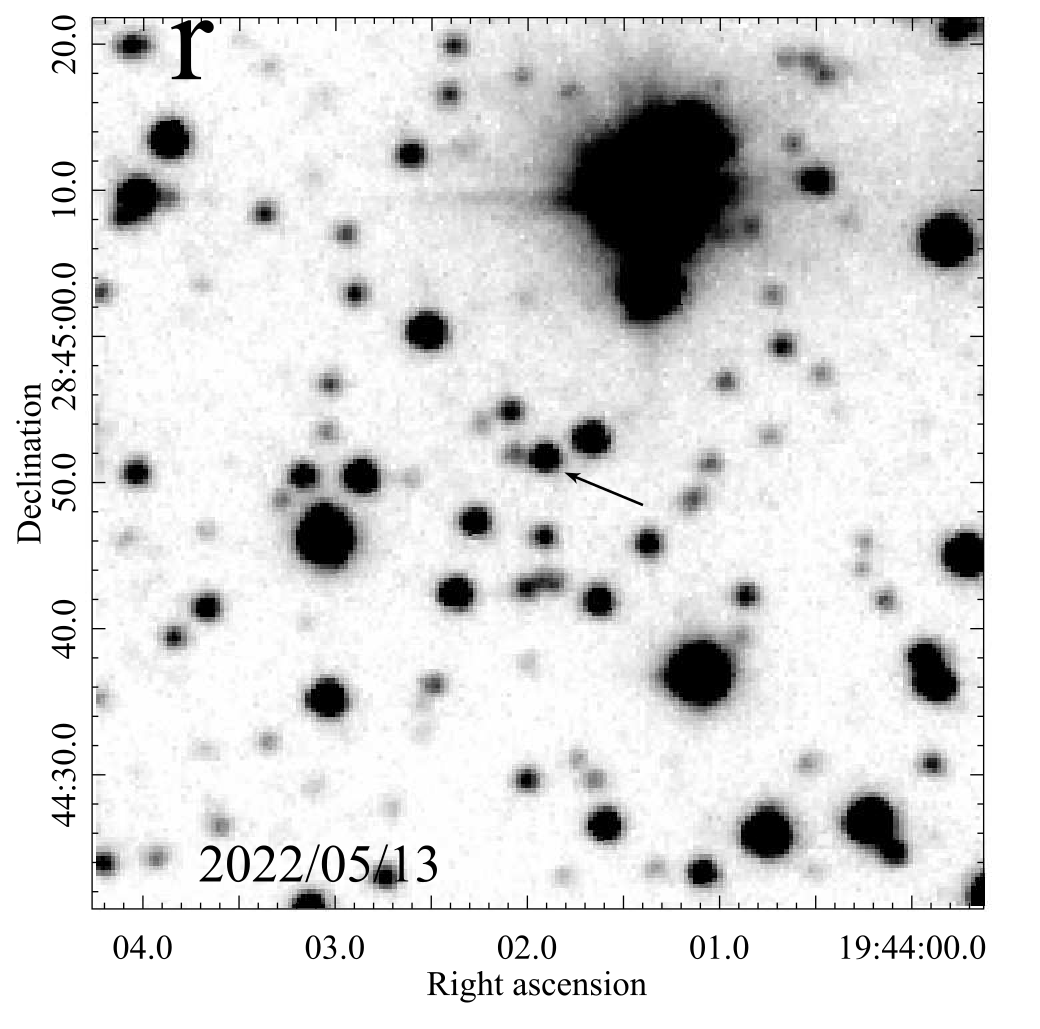}
\includegraphics[width=0.48\linewidth,clip]{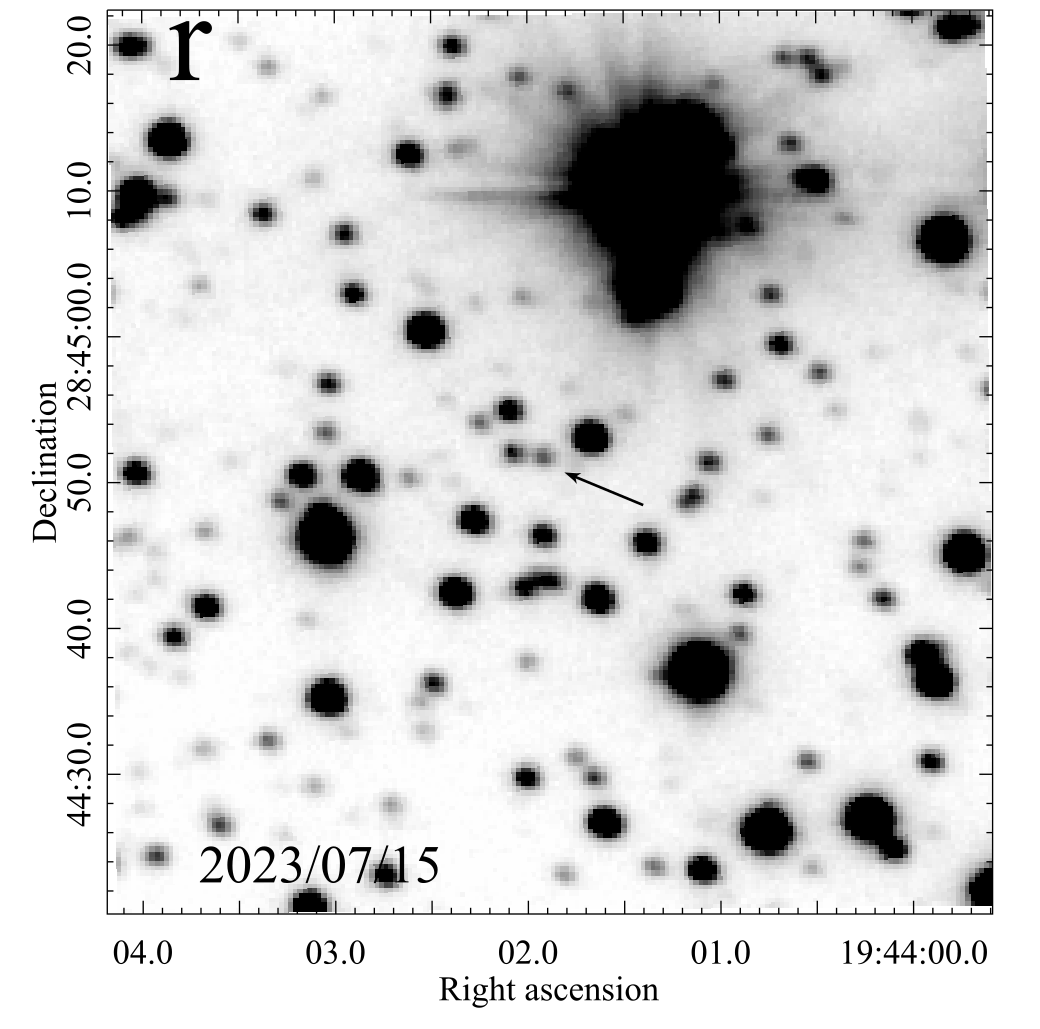}}\\
\center{\includegraphics[width=0.48\linewidth,clip]{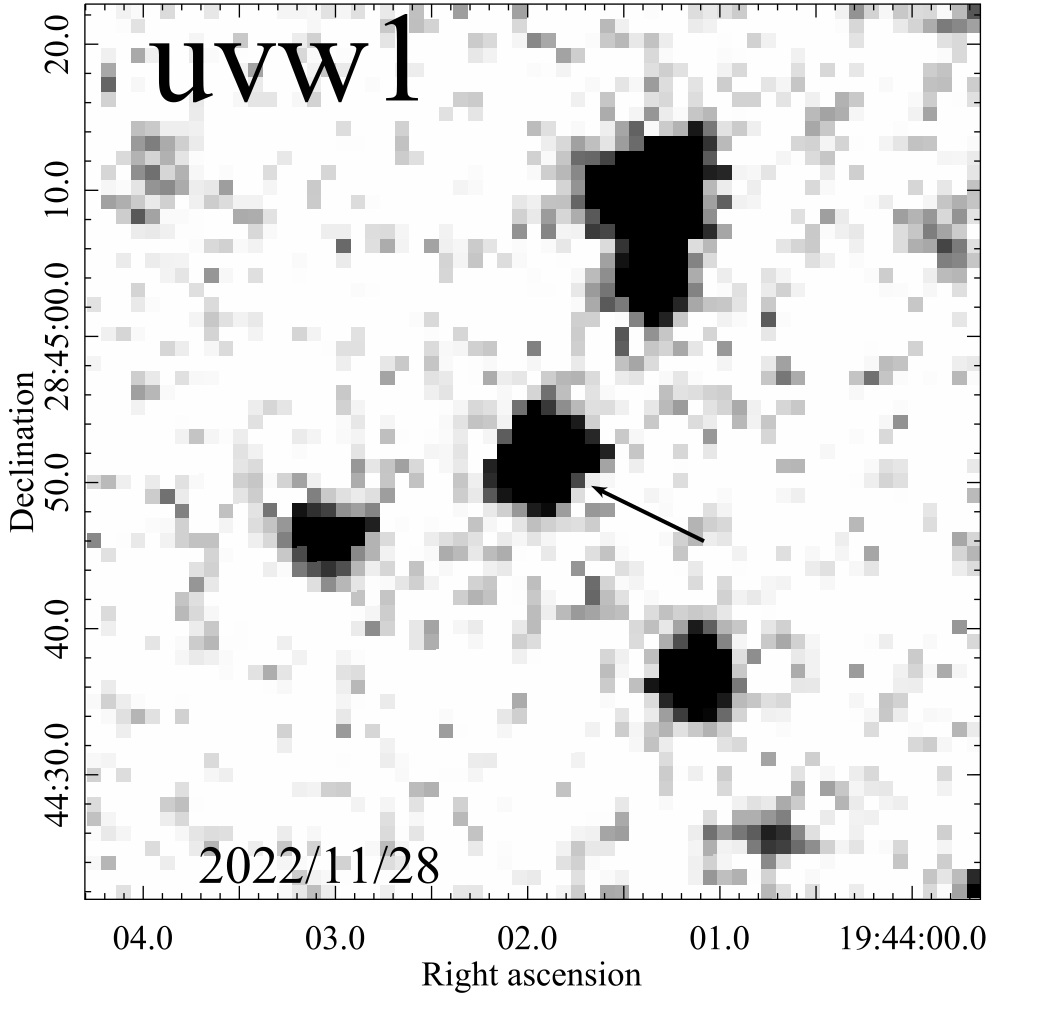}
\includegraphics[width=0.48\linewidth,clip]{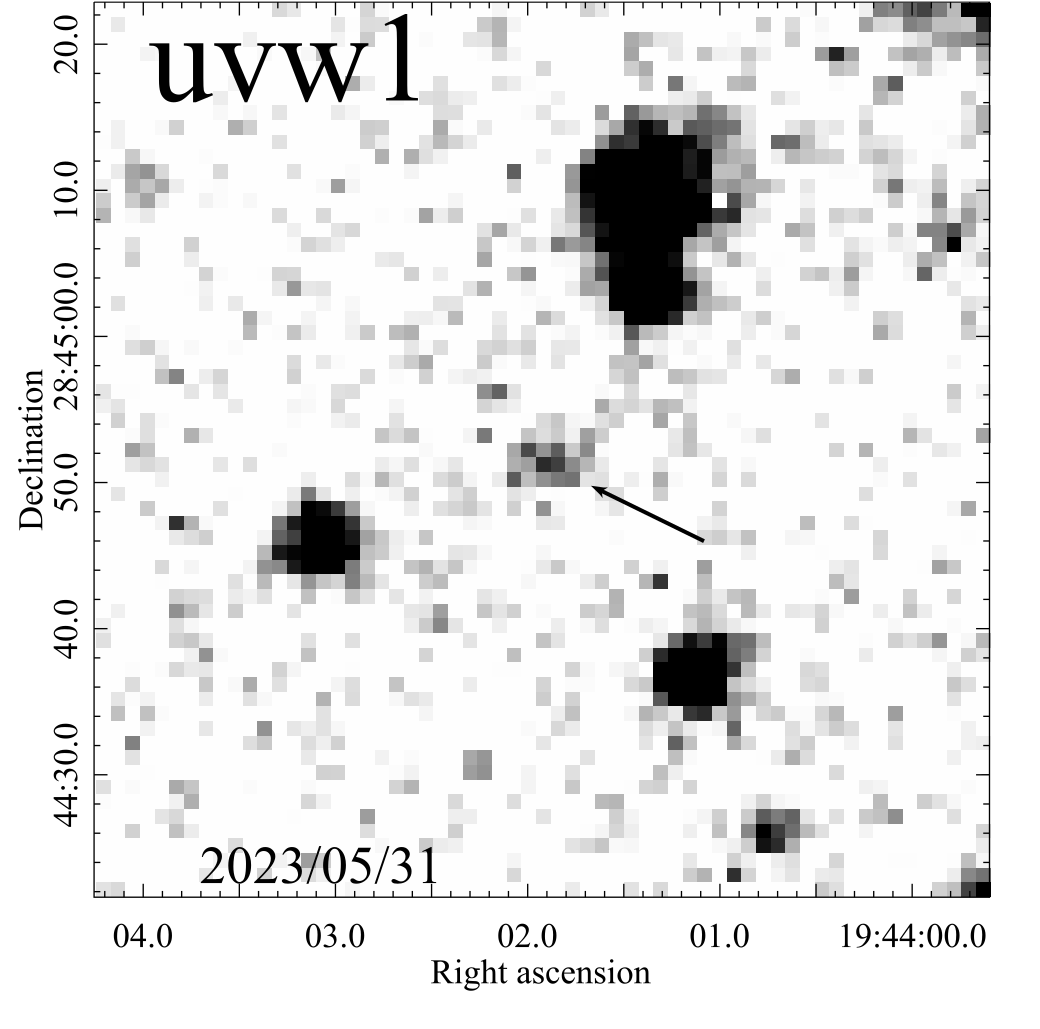}}
\end{minipage}
\caption{1\amin\ $\times$ 1\amin\ images of the J1944 field obtained by the RTT-150 in the $r$ band (top) and \sw/UVOT in the $uvw1$ band (bottom). The arrow marks the J1944 position.
The left and right panels show the high and low state of the source, respectively.
} 
\label{fig:opt-uv}
\end{figure*}

\begin{figure}
\begin{minipage}[h]{1.\linewidth}
\center{\includegraphics[width=1.0\linewidth,clip]{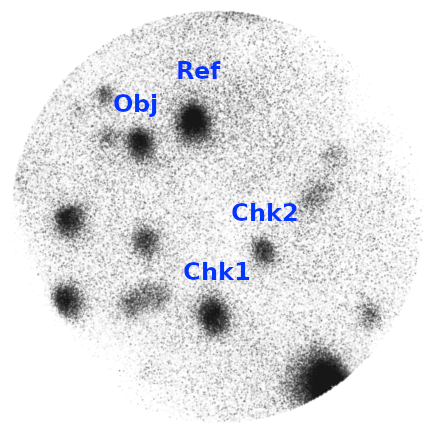}}
\end{minipage}
\caption{Image of the J1944 field with a diameter of 20\asec\ obtained with the MPPP multimode photopolarimeter. 
The source J1944 is labeled as ``Obj'', the reference star is ``Ref'', and the two check stars are ``Chk1'' and ``Chk2''.
}
\label{fig:MANIA_5}
\end{figure}

Images of the J1944 field obtained by the RTT-150 telescope in the $r$ band and \sw/UVOT in the $uvw1$ band are shown in Fig.~\ref{fig:opt-uv}. As can be seen, in 2022 the source was brighter than in 2023, both in the optical and UV, which indicates its strong variability. Fig.~\ref{fig:MANIA_5} shows an image of the J1944 vicinity 
detected by the red CSD with an exposure of 8000 sec on July 6/7, 2022.

\subsection{Variability  on long time scales}

\begin{figure*}
  \centering
  \includegraphics[width=2.0\columnwidth]{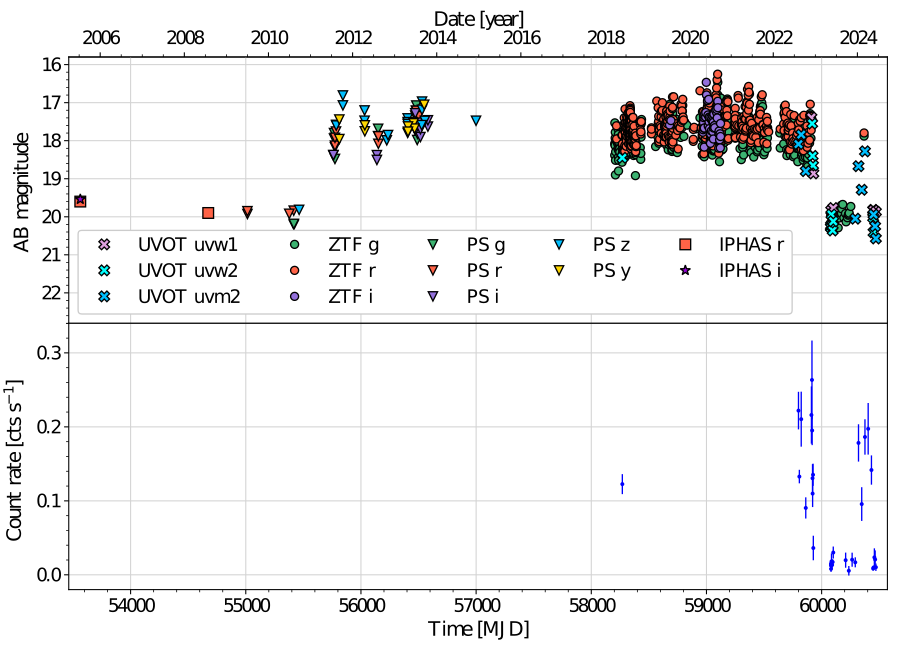}
  \put(-430,49){\includegraphics[width=1.07\columnwidth]{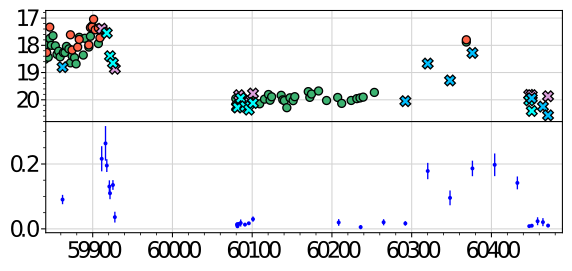}} 
\caption{Top: Optical light curves of J1944 based on the data obtained from various catalogs and instruments in the optical and UV filters notified in the legend (PS $\equiv$ Pan-STARRS).
Bottom: X-ray light curve of J1944 obtained with the \sw/XRT telescope in the 0.3 -- 10 keV range.
The recent time interval is zoomed in  and shown in the insert.
}
\label{fig:lc-dif-cat}
\end{figure*} 


To consider the optical and UV variability of the counterpart candidate of J1944 and the source itself in X-rays, we collected available archival data covering the period from 2005 to 2024. 
The light curves compiled from ZTF, Pan-STARRS, IPHAS, \sw\ UVOT and XRT data are shown in Fig.~\ref{fig:lc-dif-cat}. As seen from the figure, in the optical and UV, the source exhibits strong variability on time scales from hours to months and years. The high and low states are clearly resolved with average magnitudes of $\sim 18$\m\ and $\sim 20$\m, respectively. Transitions from the low to the high state occurred around MJD 55800 and 60300 and from the high to the low state around MJD 60000 and 60447. In the high state, the variability amplitude is about two magnitudes, and in the low state it is significantly smaller.
\begin{table*}
   \caption{List of \sw/UVOT observations of J1944 and photometry results (given in the AB system). 
   Uncertainties do not take into account 
   the systematic error of 0.03$\magup$.
      }
    \label{tab:swift-uvot}
   \centering
   \begin{tabular}{cccccc}
   \hline
   ObsID       & Date            & $uvw1$         & $uvw2$         & $uvm2$         & $u$  \\
   \hline
   00010695002 & 31-05-2018     &                & 18.45$\pm$0.04 &                &   \\
   \hline
   03111808001 & 07-08-2022 &                &                & 18.09$\pm$0.05 & \\ 
   03111808003 & 31-08-2022 &                &                & 17.84$\pm$0.07 & \\  
   03111808004 & 10-10-2022 &                &                & 18.80$\pm$0.07 & \\    
   03111808005 & 28-11-2022  & 17.39$\pm$0.05 &                &                & \\    
   03111808006 & 03-12-2022 &                &                &                & 16.62$\pm$0.04\\ 
   03111808007 & 04-12-2022 &                & 17.55$\pm$0.05 &                & \\  
   03111808008 & 07-12-2022 &                &                &                & 17.47$\pm$0.05\\  
   03111808009 & 08-12-2022 &                & 18.40$\pm$0.05 &                & \\  
   03111808010 & 12-12-2022 &                & 18.65$\pm$0.06 &                & \\  
   03111808011 & 14-12-2022 & 18.86$\pm$0.11 &                &                & \\  
   \hline
   00010695003 & 17-05-2023     &                & 20.27$\pm$0.10 &                &   \\
   00010695004 & 17-05-2023     &                & 20.23$\pm$0.10 &                &   \\
   00010695005 & 17-05-2023     &                & 20.28$\pm$0.10 &                &   \\
   00010695007 & 19-05-2023     & 19.83$\pm$0.19 &                &                &   \\
   00010695008 & 21-05-2023     &                & 19.94$\pm$0.14 &                & \\    
   00010695009 & 26-05-2023     & 20.05$\pm$0.24 & 20.24$\pm$0.18 &                & 18.98$\pm$0.17\\
   00010695010 & 31-05-2023     & 20.11$\pm$0.22 & 20.36$\pm$0.15 &                & 18.85$\pm$0.14\\
   00010695011 & 06-06-2023    & 19.77$\pm$0.17 & 20.12$\pm$0.12 &                & \\ 
   00010695012 & 21-09-2023&                &                &                & 19.12$\pm$0.12\\ 
   00010695013 & 19-10-2023 &                &                &                & 19.18$\pm$0.13\\ 
   00010695014 & 16-11-2023  &                &                &                & 18.97$\pm$0.09\\    
   00010695015 & 14-12-2023 &                &                & 20.05$\pm$0.18 & \\ 
   \hline
   00010695016 & 11-01-2024  &                &                & 18.67$\pm$0.08 & \\ 
   00010695017 & 08-02-2024 &                &                & 19.29$\pm$0.13 & \\ 
   00010695018 & 07-03-2024   &                &                & 18.28$\pm$0.06 & \\  
   00010695020 & 02-05-2024     &                &                &                & 17.33$\pm$0.03\\ 
   00010695022 & 17-05-2024     & 19.82$\pm$0.26 & 20.02$\pm$0.26 &                & 18.79$\pm$0.13\\    
   00010695023 & 20-05-2024     &  19.83$\pm$0.15  &    20.41$\pm$0.20             & 19.94$\pm$0.17  &  18.80$\pm$0.09  \\   
   00010695025 & 27-05-2024     &     &    20.25$\pm$0.19               &    &    \\  
   00010695026 & 03-06-2024     &     &                  &     &  18.58$\pm$0.11  \\  
   00010695027 & 09-06-2024     &  19.87$\pm$0.22  &            20.57$\pm$0.20       &   &    \\     
   \hline
   \end{tabular}
\end{table*}

In X-rays, J1944 also shows transitions between the high and low states, consistent in time with the transitions in the optical/UV. This  proves that the optical source is 
indeed the counterpart of J1944. 

Such variability is typical for tight binary systems with accreting compact objects, such as CVs \citep{Honeycutt&Kafka,mason&santana,11IPs-lstate}, transitional millisecond pulsars (tMSPs, \citealt{tMSPs}), or low-mass X-ray binaries \citep{mitsuda1989}.


\subsection{Optical variability on 
short time scales}

\begin{figure}
  \centering
	\includegraphics[width=\columnwidth]{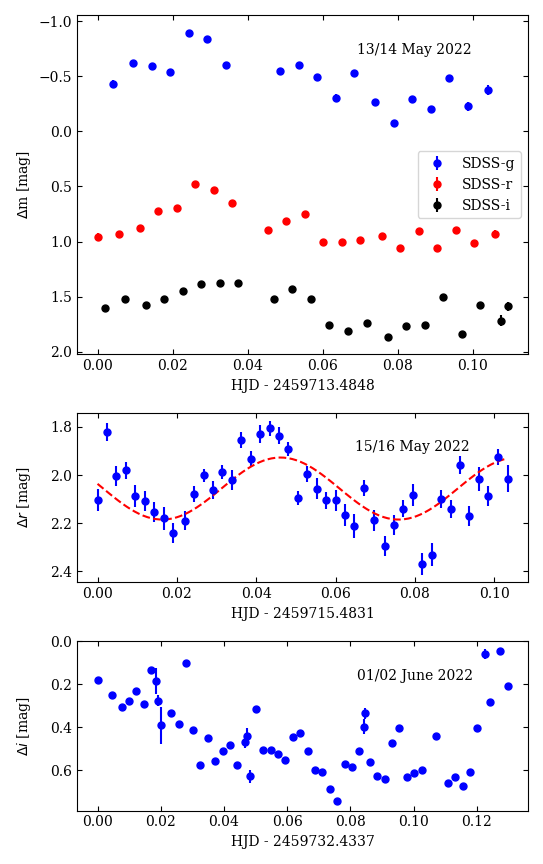}
\caption{Light curves of the optical companion candidate of J1944, obtained with the RTT-150 on May 13/14, May 15/16, June 01/02, 2022. The light curve of May 15/16 is approximated by a sinusoid with a 86~min period.
}
\label{fig:lcs_RTT}
\end{figure}

Light curves of the J1944 optical counterpart in the high state, obtained with the RTT-150 telescope, are shown in Fig. \ref{fig:lcs_RTT}. On May 13/14 2022, we carried out multiband photometric observations in the $g$, $r$ and $i$ filters of the SDSS system. All bands exhibit quasi-simultaneous brightness variations with an amplitude of $\Delta g \approx \Delta r \approx \Delta i \approx 0.5$\m\, on a time scale of $\sim 1$~ hour. Changes in brightness between individual exposures are also noticeable, significantly exceeding photometric errors $\delta m = 0.02-0.05$\m.

On May 15/16 we observed quasi-periodic brightness changes with an amplitude of $\Delta r\approx 0.4$\m\ and a variability time of 86 minutes (Fig.~\ref{fig:lcs_RTT}). In the observations of June 01/02 2022, no periodicity appears. The brightness of the system gradually decreased by $\Delta i \approx 0.4$\m\ over $\approx 0.12$~day, but then abruptly increased to the initial value in $\approx 0.01$~day.

\begin{figure*}
  \centering
	\includegraphics[width=\textwidth]{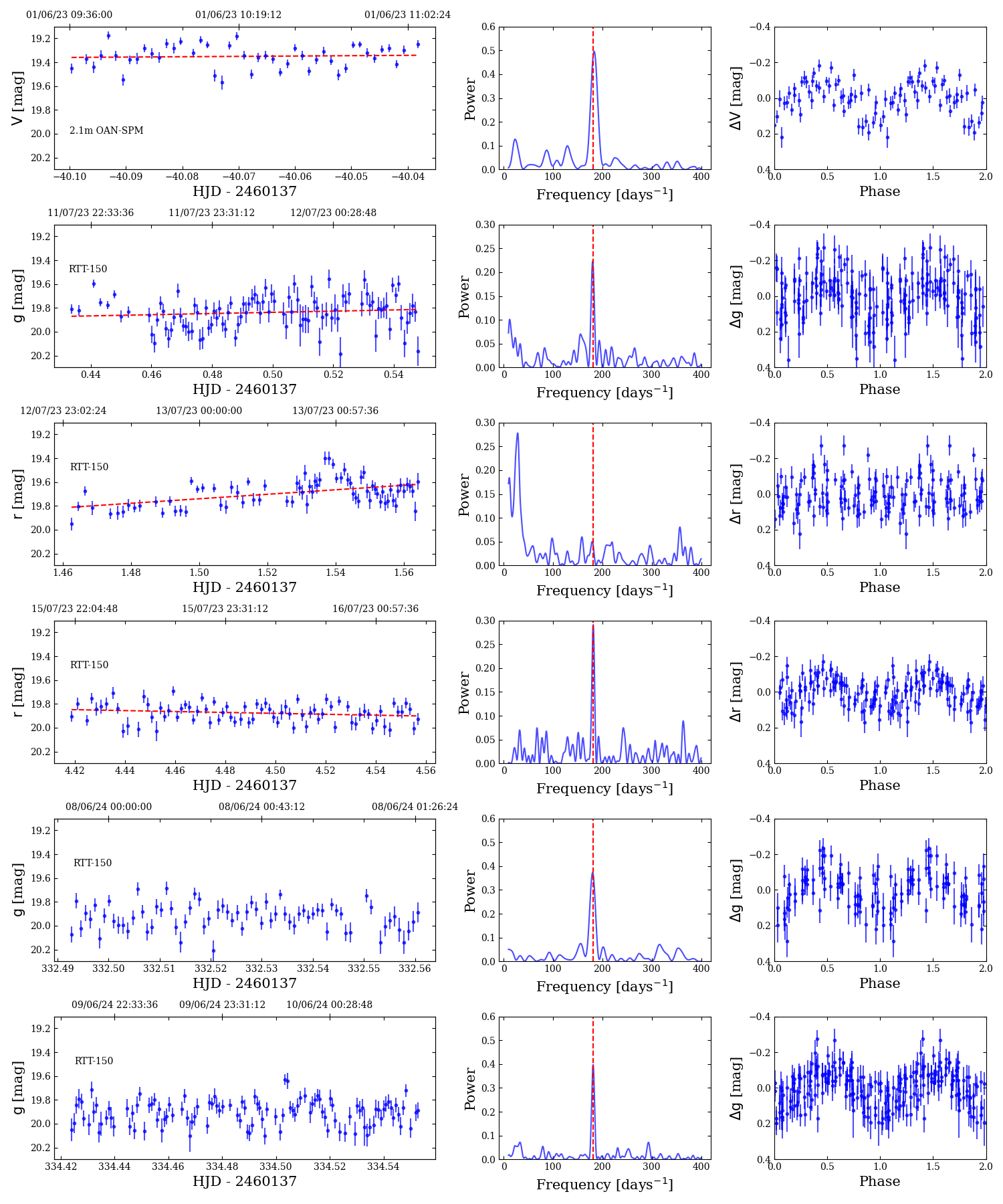}
\caption{Light curves of J1944 in the low state (left), Lomb-Scargle periodograms (middle) and  light curves folded  with the period $P_s=7.96\pm0.07$~min ($f_s = 181.00 \pm 1.66$~ day$^{-1}$) obtained from the periodograms (right). In the left plots, the dotted line shows the trend subtracted before calculating the periodograms. Data points of the OAN-SPM telescope are given in the Vega system.
}
\label{fig:periodogramms_rtt}
\end{figure*}

The J1944 light curves in the low state, obtained with the OAN-SPM and RTT-150 telescopes in 2023, are presented in Fig.~\ref{fig:periodogramms_rtt} on the left. The source does not show such strong variations as in the high state. Having performed periodogram analysis using the Lomb-Scargle method, we found a period $P_s$ equal to $7.96\pm0.07$~min (central panels in Fig. ~\ref{fig:periodogramms_rtt}). The light curves folded with this period show one peak per period with a brightness variation of no more than 0.4$\magup$ (Fig.~\ref{fig:periodogramms_rtt} on the right). To check the stability of this periodicity, additional photometric data were obtained with the RTT-150 in the $g$ filter on 07/08 and 09/10 June 2024 when the source was again in the low state. Analysis of the light curves using the Lomb-Scargle method resulted in a period of $7.948\pm0.011$~ min with an amplitude of 0.4$\magup$, which confirms the high stability of the period and amplitude of the periodic brightness variations in the low state over one year.

\begin{figure}
\begin{minipage}[h]{1.\linewidth}
\center{\includegraphics[width=1.0\linewidth,clip]{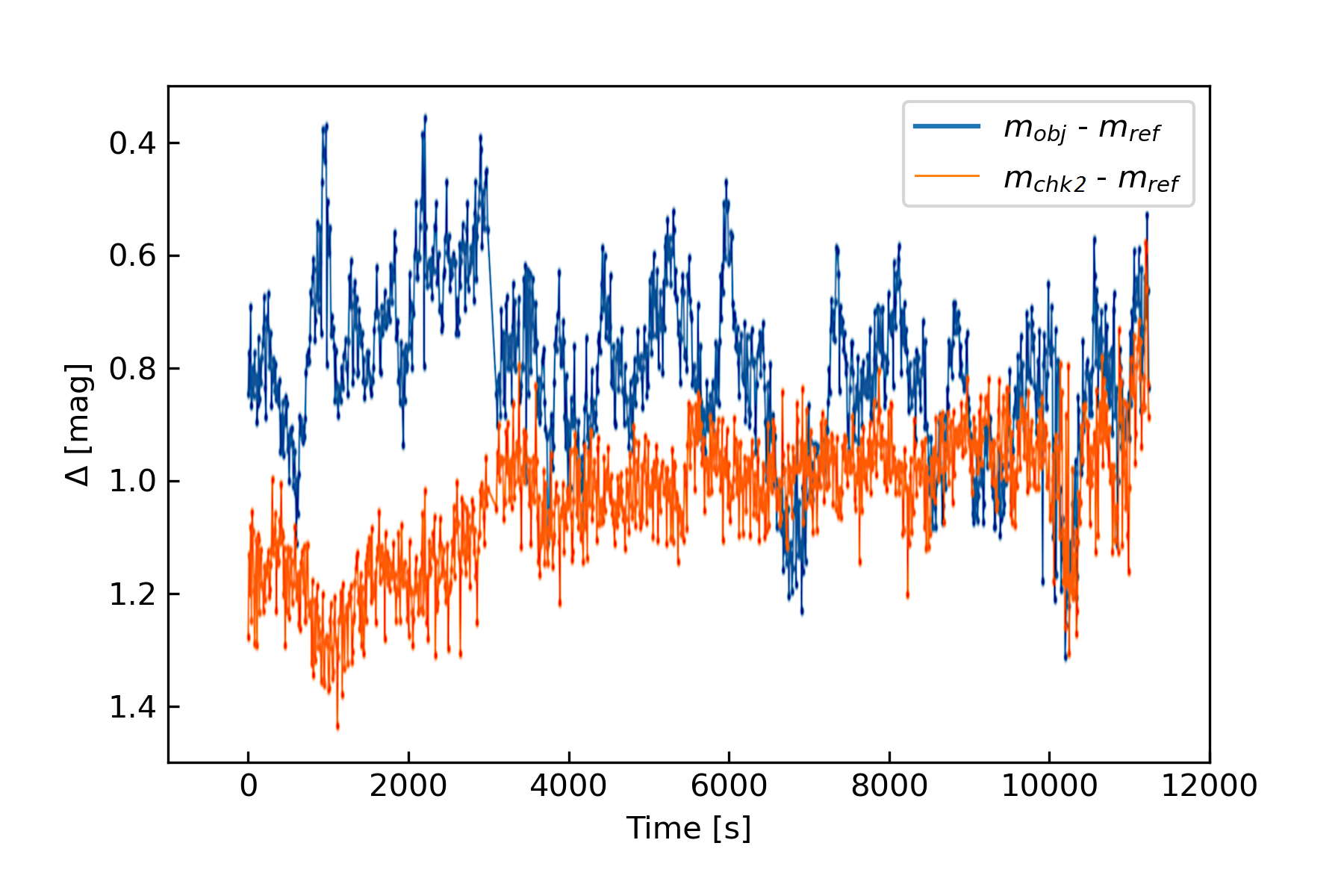}}
\end{minipage}
\caption{10-sec time resolution light curves obtained with the MPPP instrument for J1944 (blue line) and the check star (orange line).
}
\label{fig:MANIA_6}
\end{figure}

\begin{figure}
\begin{minipage}[h]{1.\linewidth}
\center{\includegraphics[width=1.0\linewidth,clip]{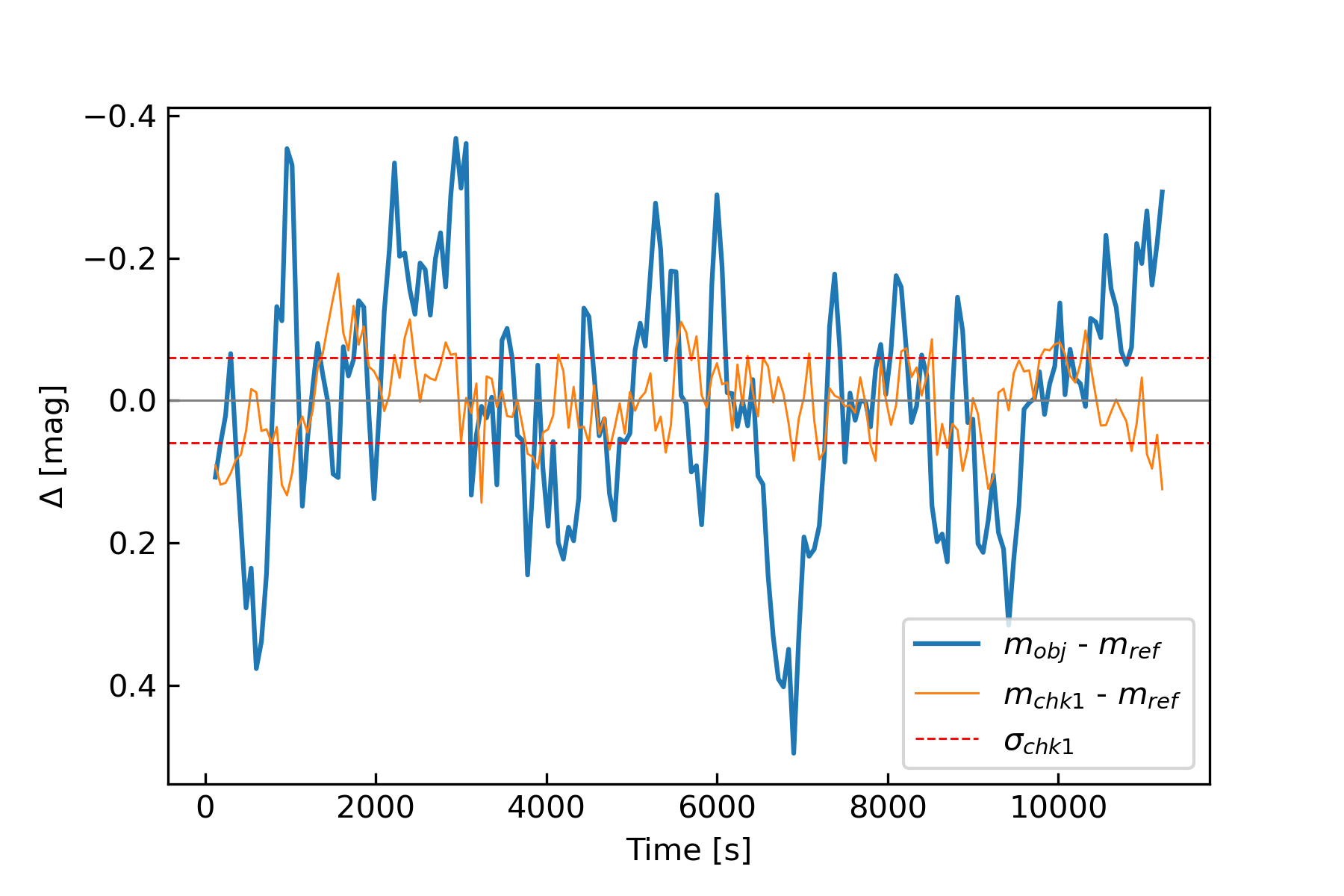}}
\end{minipage}
\caption{Light curves of J1944 and the reference star with 60-sec time resolution, obtained from observations with the multimode photopolarimeter MPPP. The red dashed lines show the 1$\sigma$ uncertainty calculated from the reference star.
}
\label{fig:MANIA_7}
\end{figure}

Based on the results of data processing from the MANIA complex (Fig. ~\ref{fig:MANIA_5}), obtained on July 6/7, 2022 at the 6-m BTA telescope when the source was in the high state, 
we constructed light curves of J1944 and field stars with a time resolution of 1, 5 and 10 sec. Fig.~\ref{fig:MANIA_6} shows the light curves with a time window of 10 sec as the difference between the instrumental magnitudes of the object and the reference star (Ref) (blue line), as well as the check star (Chk2) and Ref (orange line). The identical long-term (over hours) brightness variations are due to the effect of field vignetting and imperfect field rotation correction. As a result of reduction, light curves with a resolution of 1 min were obtained in the relative magnitudes Obj -- Ref and Chk1 -- Ref, shown in Fig.~\ref{fig:MANIA_7}.
Based on the analysis of the results, we came to the following conclusions:
\begin{itemize}
    \item no significant variability was detected on time scales of $\sim 1-10$ sec. Taking into account statistical and systematic errors, the limit on the amplitude of such variability is $\sim 0.25$\m.
    \item stochastic variations were detected on time scales of 1 -- 15 min and observed in light curves with amplitudes of 0.2 -- 0.6\m.
\end{itemize}
Apparently, this behavior  is caused by large-scale turbulence of the accretion structure around the compact object.

\subsection{Optical spectra}

\begin{figure*}
  \centering
	\includegraphics[width=\textwidth]{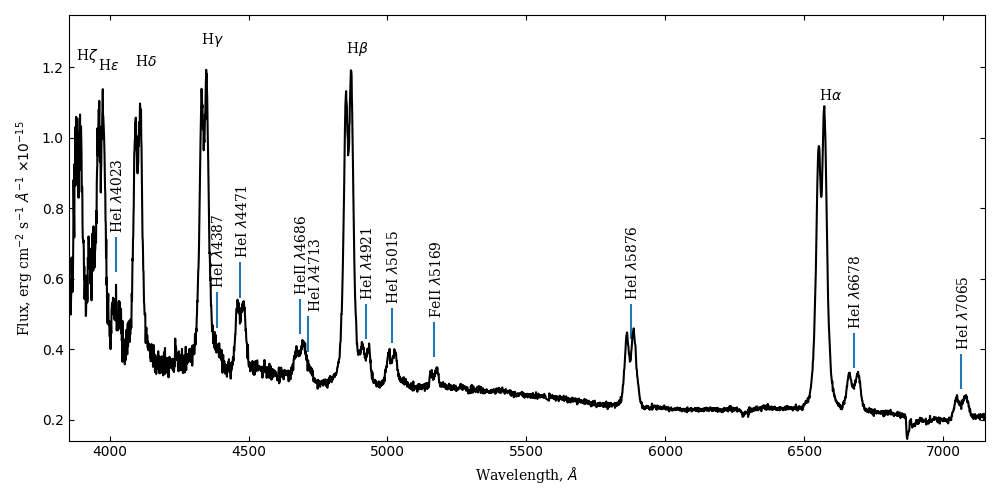}
\caption{Averaged optical spectrum of the J1944 counterpart candidate.}
\label{fig:spec_aver}
\end{figure*}

Averaged optical spectrum of   J1944  in the high state is shown in Fig. \ref{fig:spec_aver}. It reveals a blue continuum, double-peaked hydrogen emission lines of the Balmer series (H$\alpha$, H$\beta$, H$\gamma$, H$\delta$, H$\epsilon$, H$\zeta$), neutral helium (HeI $\lambda$4023, $\lambda$4387, $\lambda$4471, $\lambda$4713, $\lambda$4921, $\lambda$5015, $\lambda$5876, $\lambda$6678, $\lambda$7065), the lines of ionized helium HeII~$\lambda$4686 and iron FeII~$\lambda$5169. This is  typical for spectra of  accretion disks in  close binary systems.

\begin{figure*}
  \centering
	\includegraphics[width=\textwidth]{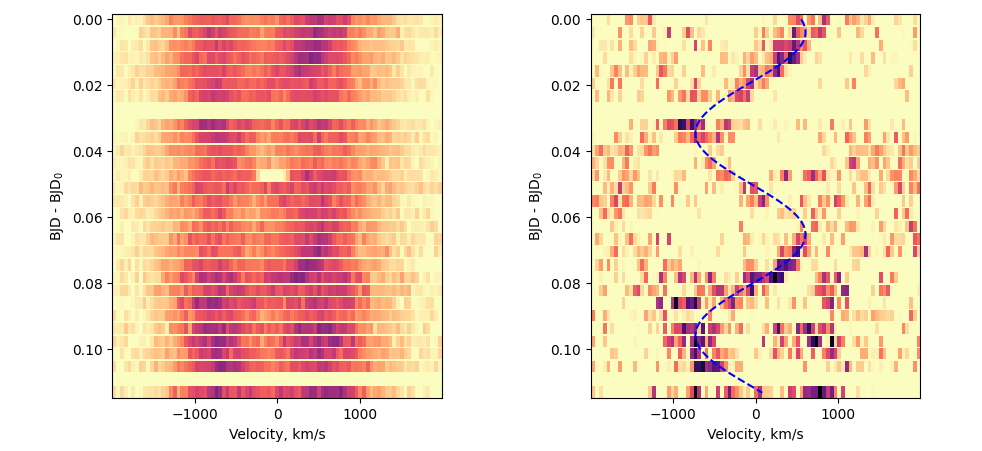}
\caption{Dynamic spectrum of the HeI~$\lambda$5876 line (left) and the residual spectrum after subtracting the average line profile (right). The initial epoch $\mathrm{BJD_0}$ corresponds to the middle of the exposure of the first spectrum.
}
\label{fig:dynam}
\end{figure*}

The dynamic spectrum of the HeI~$\lambda$5876 line, obtained from the observations on July 04/05, 2022, is shown in the left panel of Fig. \ref{fig:dynam}. It exhibits a weak S-wave, which is often presented in  spectra of CVs with accretion disks. The footprints of this wave also appear in the other lines profiles, but it is most obvious in the HeI~$\lambda$5876 line. Probably, it forms in a hot spot when the accretion flow interacts with the outer edge of the accretion disk. To separate the S-wave, we subtracted the averaged profile from the line profiles. The dynamic spectrum of HeI~$\lambda$5876 obtained from the residual profiles is shown in the right panel of Fig. \ref{fig:dynam}. 
Fitting the S-wave with a sinusoide, we estimated the radial velocity semi-amplitude of the hot spot $K_s = 650 \pm 23$~km~s$^{-1}$ and the period $P = 89 \pm 1$~min. Since the radial velocities of the hot spot are modulated by the orbital motion, the period can be considered as the orbital one.

\begin{figure}
    \includegraphics[width=\columnwidth]{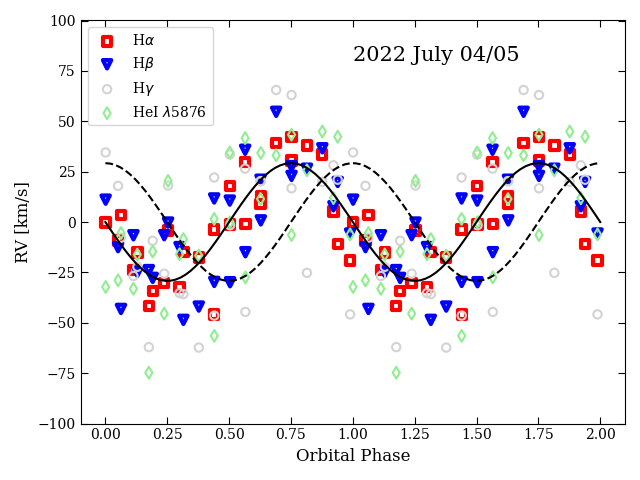}
  \hfill
    \includegraphics[width=\columnwidth]{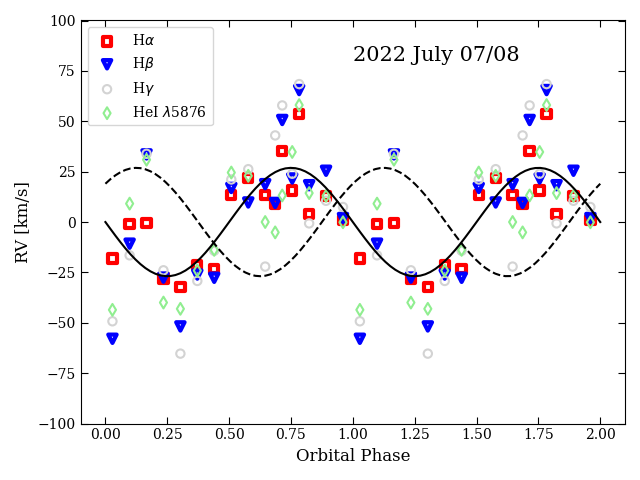}
  \caption{Radial velocity curves of the H$\alpha$, H$\beta$, H$\gamma$ and HeI~$\lambda$5876 lines wings, obtained from the observational data on July 04/05, 2022 (top panel) and July 07/08 2022 (bottom panel). The black line shows the approximating sinusoid. The dotted line corresponds to the radial velocity curve of the hot spot divided by $K_s/K_1$.
  }
\label{fig:rvs}
\end{figure}

To estimate the parameters of the system, the semi-amplitude of the radial velocity curve of the accretor $K_1$ is required. Although there is no evidence of an accretor in the spectra of J1944, its orbital motion is reflected by emission lines formed in the accretion disk. An accurate calculation of $K_1$ requires measuring radial velocities from the wings of spectral lines. The wings form closer to the accretor and are less distorted by hot spot radiation and tidal deformations of the accretion disk. To solve this problem, we used the method proposed by \cite{shafter83}. It consists of a convolution of the form
\begin{equation}
C(v_0) = \int k(v - v_0) r(v) dv,
\end{equation}
where $r(v)$ is the spectral line profile in the velocity scale, and the kernel $k$ is the difference of two Gaussians separated by a distance $a$, i.e.
\begin{equation}
\begin{aligned}
k(v) = &\exp\Bigg[-\frac{(v+a/2)^2}{2\sigma^2}\Bigg]- \\ 
          - &\exp\Bigg[-\frac{(v-a/2)^2}{2\sigma^2}\Bigg].
\end{aligned}
\end{equation}
 $a$ is selected to provide  the locations  of Gaussians  in the line wings. The standard deviations $\sigma$ must be small enough that the Gaussians do not take into account the central parts of the line. On the other hand, Gaussians must be wide enough to collect sufficient signal. The radial velocity of the line wings $v_0$ is determined from the requirement $C(v_0) = 0$. The resulting radial velocity curves were approximated by sinusoids $v_0(t) = \gamma - K_1\sin[2\pi (t/P_{orb} - \varphi_0)]$, where $\gamma$ is the average radial velocity, $K_1 $ -- the radial velocity semi-amplitude of accretor ($K_1>0$) and $\varphi_0$ -- the initial phase. To select the separation of the Gaussians $a$, we used diagnostic diagrams representing the dependences of the semi-amplitude $K$ and its error $\Delta K$ on the separation $a$. To select the optimal separation of the Gaussians $a$, we used diagnostic diagrams representing the dependences of the semi-amplitude $K$ and the error $\Delta K$ on the distance $a$. We select the distance $a=2500$~km/s as the maximum value, after that the error $\Delta K$ increases rapidly. The standard deviation is assumed to be $\sigma = 150$~km~s$^{-1}$. The resulting radial velocity curves for the lines H$\alpha$, H$\beta$, H$\gamma$, HeI~$\lambda5876$ (with the average velocity $\gamma$ subtracted) are shown in Fig. \ref{fig:rvs}. The radial velocity mean errors are $\approx 5$~km~s$^{-1}$ for the H$\alpha$ line, $\approx 9$~km~s$^{-1}$ for the H$\beta$ line, $\approx 16$~km~s$^{-1}$ for the lines H$\gamma$ and HeI~$\lambda5876$. This corresponds to the previously found orbital period. The radial velocity semi-amplitude estimate is $K_1 = 30\pm 5$~km~s$^{-1}$, and the derived mass function is $f_2(M) = 0.00017(9) M_{\odot}$.


\begin{figure*}
  \centering
	\includegraphics[width=\textwidth]{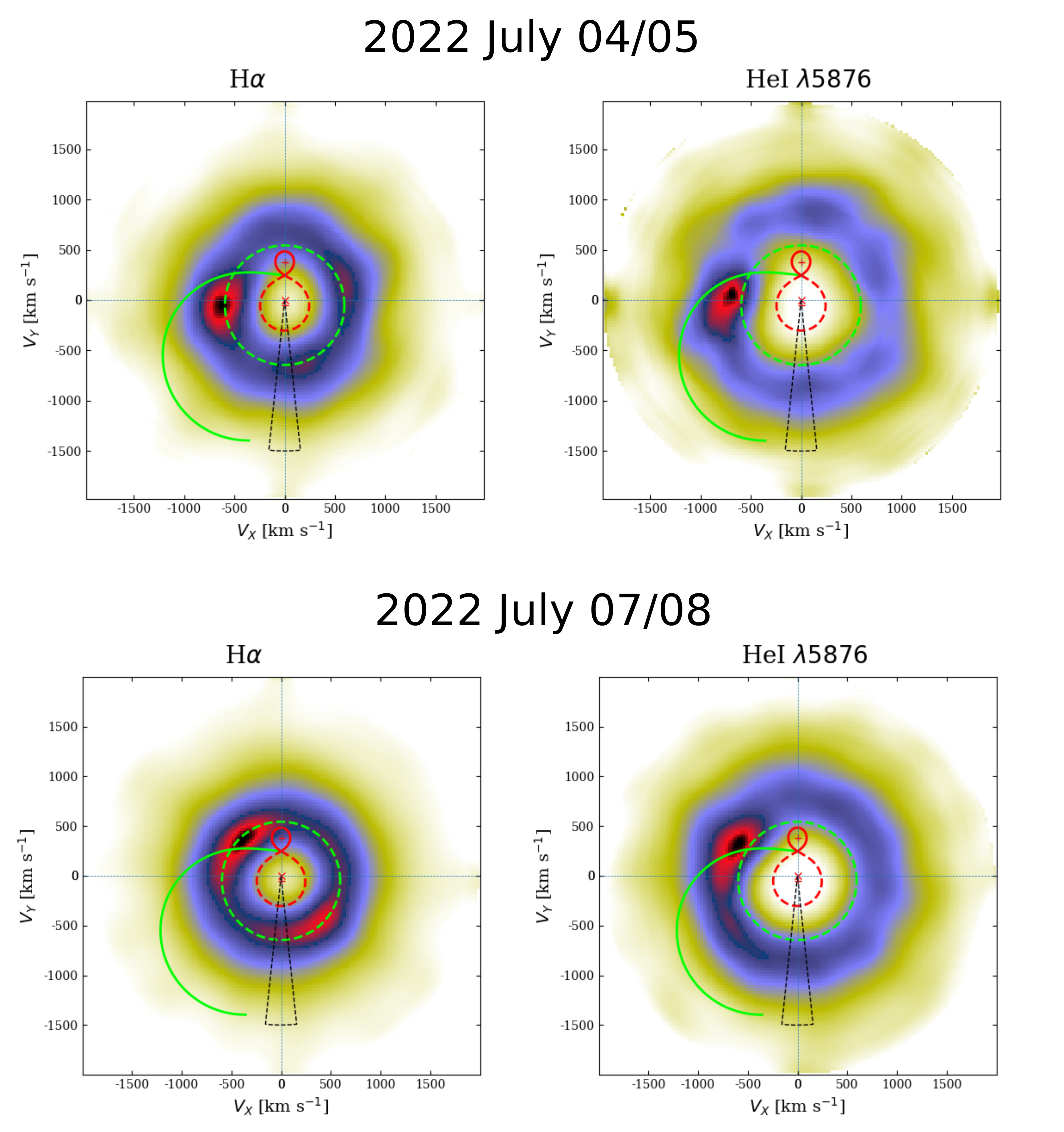}
\caption{Doppler tomograms of J1944 in the H$\alpha$ (left) and HeI~$\lambda$5876 (right) emission lines. The tomograms were reconstructed from the observational data on July 04/05, 2022 (top) and July 07/08, 2022 (bottom). The tomograms show the velocities of the Roche lobe of the primary (dotted red line) and secondary components (solid red line). There are also plotted  the ballistic trajectory (solid green line) originating from the Lagrange point L$_1$ and the particle velocities at the outer edge of the accretion disk (dashed green line) corresponding to the $3:1$ resonance. The angle of the dotted sector corresponds to the tomogram rotation uncertainty associated with the zero epoch error of the orbital ephemeris. The binary system model is calculated for the accretor mass $0.6~M_{\odot}$, mass ratio $q=0.13$ and the inclination $i=65^{\circ}$ (see ``Discussion'' section).
}
\label{fig:dopptoms}
\end{figure*}

Doppler tomograms were reconstructed from the profiles of the H$\alpha$ and HeI~$\lambda$5876 lines and are shown in Fig. \ref{fig:dopptoms}. They represent the velocity distribution of emission line sources projected onto the orbital plane. The points in this space have two coordinates: the velocity module $v$, measured from the center of mass, and the angle $\vartheta$ between the velocity vector and the line connecting the centers of mass of the stars in the system. For a more detailed introduction to the Doppler tomography method, we refer to \cite{Marsh16, Kotze15, Kotze16}. The reconstruction of Doppler maps was carried out using the code doptomog-2.0 \citep{Kotze15} with the maximum entropy method. The orbital phases of the spectral exposure midpoints were calculated from the radial velocity curves of the accretor. Due to the zero phase error $\varphi_0$, the Doppler tomograms have an uncertainty in the rotation angle of $\Delta\theta = 6^{\circ}$ and $\Delta\theta = 9^{\circ}$ in the observations of 04/05 July and 07/08 July 2022, respectively. In all tomograms presented in Fig. \ref{fig:dopptoms}, there is a ring-shaped structure that corresponds to the accretion disk. The tomograms also show a hot spot formed as a result of the accretion flow interaction with the outer edge of the accretion disk. The position of the hot spot on tomograms in the HeI $\lambda$5876 line obtained in two various nights differs by $\theta \sim 15^{\circ}$ and can be explained by initial epoch errors in orbital ephemeris. The position of the hot spot on tomograms in the H$\alpha$ line shifts significantly in $\Delta \theta \approx 45^{\circ}$ and probably reflects a change in the location of the emission source. From a comparison of tomograms in the two lines, it is obvious that HeI~$\lambda$5876 emissions are formed in higher-velocity regions of the disk that are closer to the accretor.

\subsection{X-ray spectra}

\begin{figure}
  \centering
\includegraphics[width=\columnwidth]{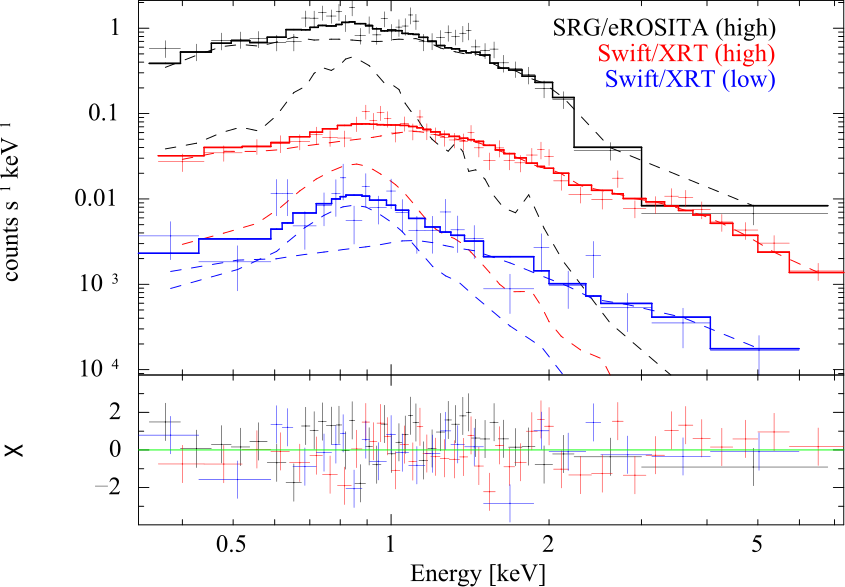}
\caption{J1944 X-ray spectra in the high and low states obtained from the \sw/XRT and SRG/\eros\ data.
The best-fitting 2-T MEKAL model (thick solid lines) and its components (dashed lines) are shown. 
Fit residuals are presented in the bottom sub-panel.
}
\label{fig:xrayspec}
\end{figure}

\begin{table}
\renewcommand{\arraystretch}{1.2}
\caption{Results of the fitting of the J1944 X-ray spectra with 
the TBABS$\times$(MEKAL+MEKAL) model. Uncertainties are at the 1$\sigma$ confidence level. The superscripts $h$ and $l$ correspond to the J1944 high and low states. VEM is a volume emission measure. $D_{415}\equiv D/$(415~pc). 
}
\label{tab:xray-fit}
\begin{center}
\begin{tabular}{lc}
\hline
\nh, cm$^{-2}$       & $5.8^{+1.3}_{-1.4} \times 10^{20}$ \\
$T_1$, keV           & $0.61^{+0.07}_{-0.09}$ \\
VEM$_1^h$, cm$^{-3}$ & $(4.1\pm0.9)\times 10^{53} D_{415}^2$ \\
VEM$_1^l$, cm$^{-3}$ & $1.5^{+0.4}_{-0.3}\times 10^{53} D_{415}^2$ \\
$T_2$, keV           & $5.94^{+1.26}_{-0.74}$ \\
VEM$_2^h$, cm$^{-3}$ & $6.1^{+0.3}_{-0.2}\times 10^{54} D_{415}^2$ \\
VEM$_2^l$, cm$^{-3}$ & $3.7^{+0.6}_{-0.7}\times 10^{53} D_{415}^2$ \\
$F_X^h$, \flux       & $5.72^{+0.28}_{-0.24} \times 10^{-12}$ \\
$F_X^l$, \flux       & $4.53^{+0.50}_{-0.46} \times 10^{-13}$ \\
$L_X^h$, \ergs       & $1.18^{+0.06}_{-0.05}\times 10^{32} D_{415}^2$ \\
$L_X^l$, \ergs       & $9.33^{+1.04}_{-0.94}\times10^{30} D_{415}^2$ \\
$\chi^2$/$\nu$       & 88.9/79 \\
$C$/$\nu$            & 26.8/20 \\
\hline
\end{tabular}
\end{center}
\end{table}

Observations with the SRG/\eros\  were performed during the high state of J1944. 872 counts from the source (excluding background) were detected in the 0.3 -- 9 keV range and they were regrouped to have at least 20 counts per energy bin. 

In the case of the \sw/XRT data, we extracted two spectra. The first one includes the data obtained on 2018, 2022 and 2024 (up to May 17) in the high state of J1944. The second one includes the data obtained on 2023 and 2024 (from May 17) in the low state of J1944. For the high state spectrum, the effective exposure  was about 8.7 ks, and 965 counts were collected from the source in the 0.3 -- 10 keV range. We regrouped it to have at least 20 counts per energy bin. For the low state spectrum, the effective exposure  was 14.1 ks, but only 130 counts were detected. We regrouped it to obtain at least 5 counts per energy bin.

Spectra extracted from the \sw\ and \eros\ data were fitted in the  0.3 -- 10 keV range using the XSPEC software package \citep{xspec}. We used the $\chi^2$ statistic for the high state data and the $C$-statistic \citep{cash} for the low state data due to the small number of counts.

We fitted the spectra with the MEKAL thermal plasma model \citep{mekal,mekal2,mekal3}. Similar models are used to describe the X-ray spectra of CVs \citep[see, for example,][]{baskill,reis,wz-sge}. At the initial analysis step, the temperature and normalization of the model were considered as free parameters for each spectrum, to evaluate possible variations of source parameters over time. To take into account the interstellar absorption, the TBABS model with the WILM elemental abundances \citep{wilm} was used. The equivalent absorbing hydrogen column density \nh\ for all spectra was linked.

For the high-state \eros\ and \sw\ spectra, the reduced $\chi^2$ are $\chi^2_\nu=1.4$ (degrees of freedom $\nu=37$) and 1.2 ($\nu=41$), and for the \sw\ low-state spectrum, $C_\nu=2.1$ ($\nu=22$). The fits show some excesses of the observed flux compared to the model at energies $\lesssim 1$ keV.

Then we used the two-temperature model 2-T MEKAL. In this case, the fits are acceptable. 
For all the three spectra,  the derived temperatures $T_1$ and $T_2$  
are consistent 
within 2$\sigma$ uncertainties. For the two high state spectra, the normalizations of the thermal components are consistent, while for the low state data the normalization is lower, as expected from the light curve (Fig.~\ref{fig:lc-dif-cat}, bottom). The parameters for the latter case  are determined poorly due to the small number of counts. Finally, we 
fitted the three spectra simultaneously, assuming that all parameters are common except for normalizations of model components. The results of the fit, the unabsorbed fluxes $F_X^h$ and $F_X^l$ in the high and low states in the 0.3 -- 10 keV range, as well as the luminosities $L_X^h$ and $L_X^l$ are presented in Table ~\ref{tab:xray-fit}. The spectra and the best-fitting model are shown in Fig.~\ref{fig:xrayspec}.


\section{Discussion}

The considered archival and original optical, UV and X-ray data indicate that J1944 is a tight binary accreting system with an orbital period of $\approx$1.5 hours. The orbital period is unambiguously determined from the analysis of time variations in the emission spectral lines of the hydrogen Balmer series, ionized and neutral helium. The double-peaked emission lines and the Doppler tomography also indicate the presence of an accretion disk and hot spots associated with the collisions of disk material with a stream of gas from the donor star overflowing its Roche lobe. Occasionally, a signature of the orbital period also appears in the optical light curves obtained from the multiband photometry and has one maximum per period. However, in most observations, the optical brightness varies stochastically on time scales ranging from minutes to hours, exhibiting sporadic flares at a level of a few tenths of a magnitude, observed simultaneously in several filters. This indicates that the accretion process is unstable. We found no evidence of eclipses in this binary system. At the same time, we discovered that occasionally, on scales of several months/years, J1944 abruptly transits into fairly long (from six months or more) high and low states, differing in optical brightness by more than one magnitude. Transitions occur simultaneously in X-rays, the optical and UV ranges, and they are likely associated with a significant change in the accretion rate. In the low state the optical light curves do not exhibit flare activity, but show regular variations in the emission intensity with a period of $\approx$8 min. In a high state the optical spectrum of the system is almost completely dominated by the disk emission. At the same time, in the low state the spectral energy distributions in the optical  and UV likely correspond to a stellar spectrum (see below). 

The mentioned observational properties of  J1944 are typical for a CV consisting of an accreting white dwarf and a non-degenerate late-type star, 
which is a donor of accreting material. The presence of an 8-min period excludes the interpretation of the object as a binary system with a neutron star, expected to be a millisecond pulsar. Assuming that J1944 is the CV, below we estimate the binary system's parameters in detail, discuss their relevance to those expected for such objects and consider a  possible CV type.


\subsection{Masses of components and orbital parameters of the CV, temperature of the white dwarf}

\begin{figure}
  \centering
	\includegraphics[width=0.8\linewidth]{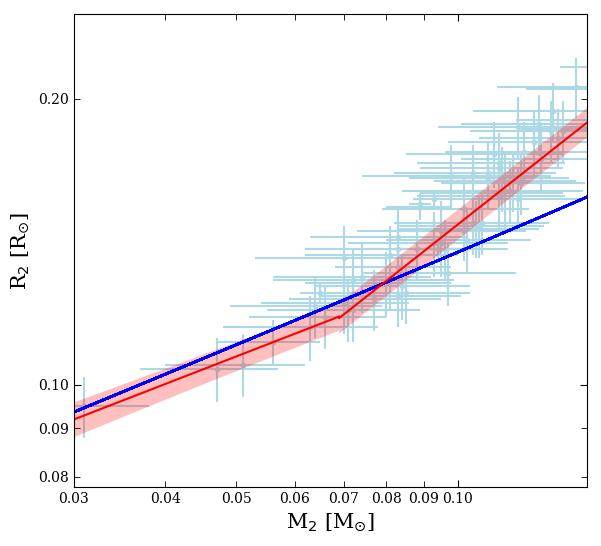}
\caption{The ``radius -- mass'' relation for the donor in the CV. Data points and error bars correspond to measurements from \cite{Knigge06}. Red lines are power-law approximations with the 1-$\sigma$ uncertainties (pink polygons) for donors with masses $M_2 < 0.069$~M$_{\odot}$ (period bouncers) and $0.069$~M$_ {\odot} < M_2< 0.20$~M$_{\odot}$ obtained by \cite{Knigge06, Knigge11}. The blue line points out the effective radii of the donor Roche lobe, calculated for the masses of the white dwarf $M_1 \in 0.20-1.44$~M$_{\odot}$.}
\label{fig:R2M2}
\end{figure}

An estimate of the secondary component (donor) mass $M_2$ can be found from the requirement that it fills the Roche lobe. According to \cite{Sirotkin10} the effective radius of the Roche lobe $R_L$ of the donor is related to the mass ratio $q=M_2/M_1$ by the expression
\begin{equation}
R_L=A\frac{0.5126 q^{0.7388}}{0.6710q^{0.7349}+\mathrm{ln}(1+q^{0.3983})},
\label{eq_rl}
\end{equation}
where $A$ is the semi-major axis of the system, related to the masses of the components and the orbital period by Kepler's third law $A=(M_2(1+1/q)P_{orb}^2)^{1/3}$, $M_1 $ --- mass of the accretor (white dwarf). Equation (\ref{eq_rl}) is a refinement of the relation from \cite{Eggleton83} for polytropic stellar models (polytropic exponent $n=3/2$), which reproduce better the fully convective donors in CVs below the orbital period gap. On the other hand, the radius of a star is evolutionarily related to its mass. We used the empirical relationship $R_2(M_2)$ presented in \cite{Knigge11} and shown in Fig. \ref{fig:R2M2}. It is approximated by two red lines. The first corresponds to donors with a mass $M_2<0.069 M_{\odot}$ and the binary system evolves with increasing period (period bouncer). The second line is an approximation for masses $M_2 \in 0.069 - 0.20 M_{\odot}$. The plot also shows the relation (\ref{eq_rl}) of the Roche lobe radius (blue line), calculated for white dwarf masses $M_1\in 0.2-1.44 M_{\odot}$ ($R_L$ weakly depends on $M_1$). From Fig. \ref{fig:R2M2} it is clear that the donor radius is consistent with the radius of the Roche lobe at $M_2 \le 0.08\pm0.01 M_{\odot}$ which is the upper limit of the mass $M_2$.

\begin{figure}
    \includegraphics[width=\columnwidth]{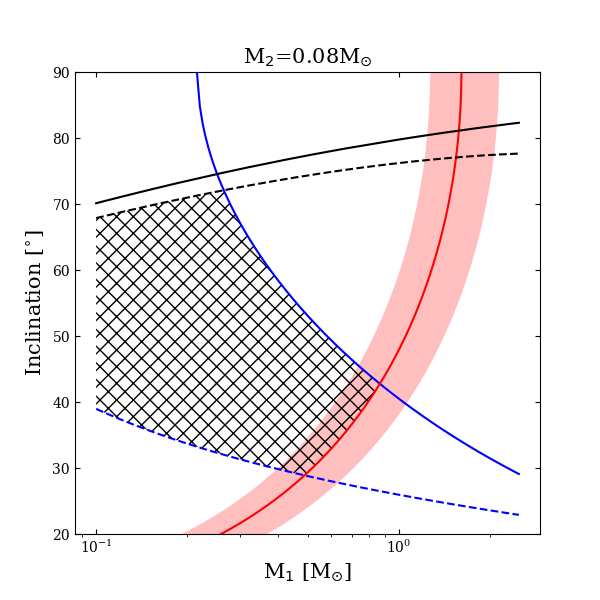}
  \hfill
    \includegraphics[width=\columnwidth]{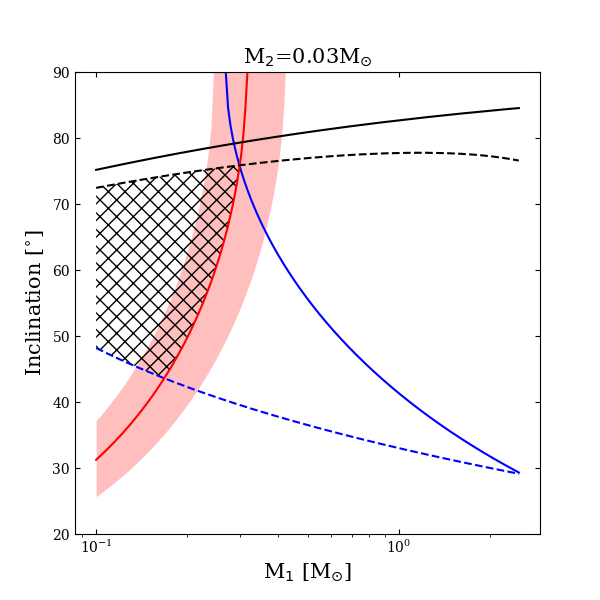}
  \caption{Set of solutions in the $i-M_1$ plane. The solutions for the donor mass $M_2 = 0.08M_{\odot}$ are shown at the top, and for $M_2 = 0.03M_{\odot}$ at the bottom. The red solid lines are solutions that are derived from the white dwarf radial velocity semi-amplitude $K_1$. Blue solid lines are solutions based on the assumption that the peaks of the H$\alpha$ lines form at the edge of the accretion disk of the maximum radius $R_{\mathrm{max}}$. Similar solutions for an accretion disk of the minimum radius $R_{\mathrm{min}}$ are marked with blue dotted lines. The black solid line shows the inclinations $i$, at which the eclipse of the accretor occurs, and the black dotted line shows the inclinations up to which there is no eclipse of the accretion disk of radius $R_{\mathrm{min} }$.}
\label{fig:solutions}
\end{figure}

Top panel of Fig. \ref{fig:solutions} shows the solution in the $i-M_1$ plane ($i$ is the inclination) corresponding to the found radial velocity semi-amplitude $K_1$ for the donor mass upper limit $0.08M_{\odot}$. A decrease in the donor mass shifts this curve to the left, which leads to a decrease in the accretor mass (bottom panel of Fig. \ref{fig:solutions}). The resulting solution for $i=90^{\circ}$ gives a limit on the accretor mass $M_1 \lesssim 1.55~M_{\odot}$ that exceeds the Chandrasekhar limit for white dwarfs.

To reduce the range of acceptable system parameters, we estimated the velocity  projection of the outer edge of the accretion disk $v_{\rm out} \sin i$. According to \cite{Yakin13, Neustroev16}, we assumed that the distance between the peaks in the emission lines (in velocity units) is equal to the velocity projection $v \sin i$ of the disk ring where the emission is formed. The spectral line intensity is sensitive to the local parameters of the accretion disk (temperature, pressure), and therefore the distance between the peaks of different lines corresponds to different rings of the disk. The minimum $v \sin i$ was calculated for the H$\alpha$ line and it is equal to $v \sin i \approx 465$~km/s. We accept this value as the estimate of $v_{\rm out} \sin i$.

Assuming that the motion of particles at the outer edge of the disk is close to Keplerian, the projection of the rotation velocity is equal to
\begin{equation}
    v_{\rm out} \sin i = \Big(\frac{GM_1}{R_{\rm out}}\Big)^{{1}/{2}} \sin i,
\end{equation}
where $R_{\rm out}$ is the radius of the accretion disk. According to \cite{Paczynski77, Warner95}, the size of the accretion disk $R_{\rm out}$ is limited by the tidal forces of the secondary component and is estimated by the formula
\begin{equation}
R_{\max} = A \frac{0.6}{1+q}.
\label{r_min}
\end{equation}
The solution corresponding to this projection of the rotation velocity for $M_2 = 0.08 M_{\odot}$ is plotted on the $i-M_1$ plane shown in the top panel of Fig. \ref{fig:solutions}. As can be seen, for a donor mass $M_2 = 0.08 M_{\odot}$ we have an accretor mass $M_1 \approx 0.88 M_{\odot}$, consistent with the Chandrasekhar limit, and an orbital inclination $i\approx 43^{\circ }$. In Fig. \ref{fig:solutions}, we also show restrictions on orbital inclination, due to the lack of eclipses in the system. The first restriction reflects the absence of an eclipse of the accretor, and the second --- the accretion disk. For the last constraint, we used the relation for the minimum radius of the accretion disk from \cite{Hessman90}
\begin{equation}
    R_{\mathrm{\min}} = 0.0859 A q^{-0.426}.
\end{equation}

The bottom panel of Fig. \ref{fig:solutions} shows a solution set for a donor with the mass $M_2 = 0.03 M_{\odot}$. The intersection of the solution curves obtained for $K_1$ and $v_{\mathrm{out}} \sin i$ corresponds to the limiting orbital inclination $i$. At lower masses $M_2$ the system parameters are inconsistent with the absence of eclipses in J1944. Therefore, $M_2 = 0.03 M_{\odot}$ can be considered as a lower limit for the donor mass. From Fig. \ref{fig:solutions} it follows that the accretor mass is in the range $M_1 \in 0.30 - 0.88 M_{\odot}$, and the orbital inclination is in the range $i\in 40-75^{\circ}$. It should be noted that the restrictions on the accretor mass and orbital inclination are derived from the assumption that the H$\alpha$ line peaks form at the edge of the accretion disk. If they are formed in a ring of radius $R\in R_{\min} - R_{\max}$, then the range of permissible parameters of the system significantly expands. The regions of feasible solutions for two limiting donor masses are shown in Fig.~\ref{fig:solutions} by shaded areas and are consistent with the values expected for CVs.

\begin{figure}[!htb]
    \centering
    \includegraphics[width=\linewidth]{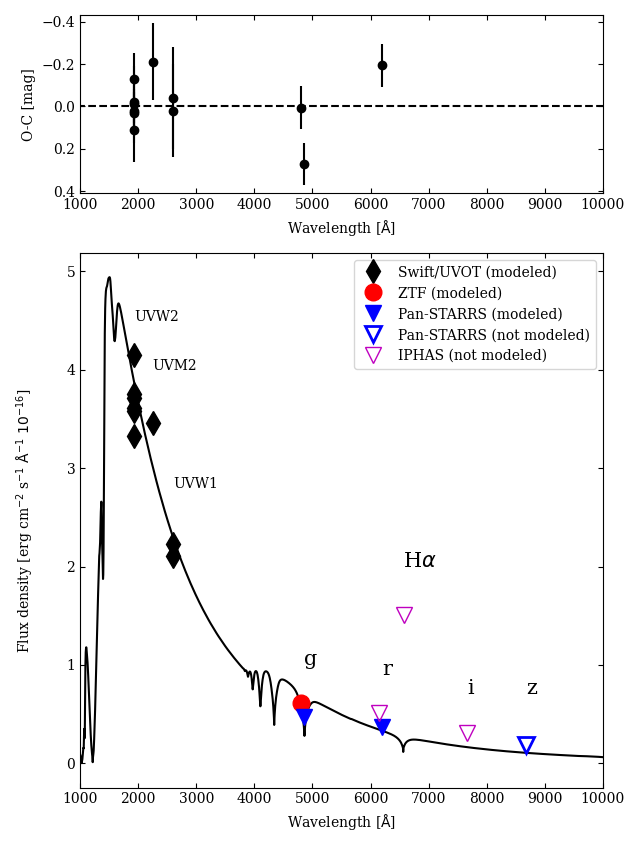}
     \label{Fig:Data1}
    \caption{The bottom panel shows the spectral energy distribution of J1944 in the low state constructed from the  photometric data and the best-fitting model of the white dwarf. Filled symbols indicate the fluxes used to estimate atmospheric parameters; empty symbols were not taken into account. The residuals of the observed and simulated fluxes are shown in the top panel. 
  }
\label{fig:WD-spec}
\end{figure}

In 2023, the source J1944 transitioned to the low state with a brightness of $\langle g \rangle \approx 20\magup$. In the optical and UV observations, the orbital variability of the object was absent. One can assume that in the low state the main source of radiation is the white dwarf. Based on that, we attempted to reproduce the spectral energy distribution of J1944 (Fig.~\ref{fig:WD-spec}) using spectra of hydrogen model atmospheres of white dwarfs calculated by \cite{Koester10}. The fit was performed using the least squares method and based on the observed fluxes in the $uvw1$, $uvm2$, $uvw2$, $g$ and $r$ filters. The flux errors in the $g$ and $r$ filters are equal to the approximate brightness variability 0.1\m\ 
of J1944 in the low state. The fluxes in the $i$ and $z$ filters were excluded due to a possible contribution of the accretion disk and donor. The flux in the $u$ filter was excluded due to contamination from nearby stars. The J1944 photometric fluxes have been corrected for the interstellar extinction using the extinction curve from \cite{Fitzpatrick99}. We used the color excess $E(B-V)=0.06\magup \pm0.04\magup$ for a distance of 415 pc in the source direction according to the interstellar extinction maps\footnote{\url{https://stilism.obspm.fr/}} \citep{extmap1,extmap3,extmap2}. The extinction for J1944 in the Johnson $V$ filter is $A_V = 3.1 E(B-V) = 0.18\magup$. The best fit 
of the spectral energy distribution was obtained for the white dwarf temperature $T=14750 \pm 1250$~K. This temperature is typical for a CV below the period gap \citep{Townsley09}. The agreement with the observed fluxes is satisfied for the white dwarf radius $R_1 = 0.0094 - 0.0124R_{\odot}$, which corresponds to the mass $M_1 = 0.61 - 0.86 M_{\odot}$ \citep{Nauenberg72}. For details of the method for modeling the spectral energy distribution, we refer to \cite{kolbin24}. The resulting estimate of $M_1$ is consistent with its limitations presented above. 

The low-mass donor is much cooler and its contribution is not evident in our data, but may be revealed in the red and infrared ranges of the spectrum. Observations in these ranges are necessary for its firm detection and classification.

As a result, our data and estimates show that the accretor is a white dwarf, and the donor is a low-mass star of a late spectral class, which confirms the CV nature of J1944.

\subsection{Variability of radiation and periodicity}

As previously noted, the J1944 can remain in the high and low states for several years. According to Fig.~\ref{fig:lc-dif-cat}, the object was in the high state at least from July 2011 to December 2014 (3.4 years), then from March 2018 to December 2022 (4.7 years) and also from January to early May 2024 (0.5 years). The actual state of the source in the period from 2015 to 2017 could be clarified by the \gaia\ data, but at the moment the catalog contains only the averaged magnitude for all the observation period. J1944 was in the low state through 2023, and possibly between July 2005 and September 2010 (approximately 5.2 years), although the data is quite sparse. Currently the source has been in the low state since May 2024. The photometric observations carried out with the RTT-150 telescope on June 15, 2024 resulted in the following magnitudes: $g$ = 19.82\m\ $\pm$ 0.03\m, $r$ = 19.82\m\ $\pm$ 0.09\m, $i$ = 19.76\m\ $\pm$ 0.09\m\ and $z$ = 19.64\m\ $\pm$ 0.07\m. These values are in a good agreement with the Pan-STARRS, ZTF, and \sw/UVOT data obtained in the low state (see Fig. \ref{fig:lc-dif-cat} and Table \ref{tab:swift-uvot}).

High and low states are observed in CV types such as polars \citep{mason&santana}, intermediate polars \citep{6-IPs-lowstate,11IPs-lstate}, and VY Scl-type nova-like systems \citep{Honeycutt&Kafka}. The presence of an accretion disk excludes 
the interpretation of the source as a polar. 

The measured orbital period of the system $P=89\pm1$ min is close to the minimum 
value of $\sim 80$ min for CVs with hydrogen donors \citep{Knigge06}. Intermediate polars \citep{Pspin-Porb} have similar periods, while VY Scl type systems have longer periods, usually 3 -- 6 hours \citep{VY-Scl2023}.

Short optical brightness variations with the period $P_s=7.96$ min, that we 
derived from the photometric observations in 2023, when J1944 was in the low state, may be associated with the rotation of the white dwarf. This period is not detected in the high state of J1944, which is apparently due to the dominance of the accretion disk.
If $P_s$ is the real spin period of the white dwarf, then the ratio of the spin period to the orbital period $P_s/P=0.09$ is close to the typical values of 0.01 -- 0.1 for intermediate polars \citep{Pspin-Porb}.

On the other hand, $P_s$ could be the pulsation period of the white dwarf (e.g., \citealt{szkody2021}). Estimates of mass and temperature indicate that our object may be in an instability strip. However, confirmation of an 8-min periodicity with a one sec accuracy from the photometric data of RTT-150 in 2024 is a strong argument in favor of the white dwarf spin. 
In addition, the observed amplitude of the 8-min brightness variability is 0.4\m. It significantly exceeds the observed amplitudes related with  pulsations of white dwarfs, typically less than  0.2\m\ (see, e.g., \citealt{Fontaine2008}).

In the high state, only stochastic variations were detected on times of 1 -- 15 min with amplitudes of 0.2 -- 0.6\m, and they were absent on smaller time scales.


\subsection{X-ray spectra}

The X-ray spectrum of J1944 can be described by a two-temperature optically thin plasma model, indicating temperature inhomogeneity of the emitting region. The derived temperatures are typical for CVs \citep[e.g.][]{baskill,worpel}.

Using the color excess $E(B-V)$ for J1944 mentioned above   and the empirical relation from \citet{foight}, we estimated the hydrogen column density along the line of sight \nh=$(5.2\pm3.8)\times10^{ 20}$ cm$^{-2}$. This is consistent with \nh\ obtained from  the X-ray spectral analysis (see Table~\ref{tab:xray-fit}).

The estimated X-ray luminosities in the high and low states are $\approx 10^{32}$ and $\approx10^{31}$ \ergs, respectively. 
The \xmm\ scan of the source was performed
on May 3, 2013, when J1944 was in the high state (see Fig.~\ref{fig:lc-dif-cat}). 
The luminosity was $\sim 2$ times greater than in \sw/XRT observation data averaged over several years. Taking into account the strong brightness variability of the source in the high state (Fig.~\ref{fig:lc-dif-cat}), this is not surprising. The resulting X-ray luminosities are consistent with those of CVs with similar orbital periods \citep{11IPs-lstate}.
The optical fluxes (calculated for the $g$ band) in the high and low states are about 10 times lower than the X-ray fluxes in the respective states. This flux ratio is also typical for CVs \citep{lapalombara}.


\subsection{Accretion rate}

Double-peaked emission lines are observed in the optical spectrum throughout the orbital period in the J1944 high state.
This indicates the presence of an accretion disk, confirmed by the Doppler tomograms. In the high state, the white dwarf and the donor star are not 
revealed in the spectrum.

In the low state, the spectral energy distribution corresponds to the stellar one (Fig.~\ref{fig:WD-spec}). However, while the object was in the low state, from the IPHAS data the flux density in the H$\alpha$ line was three times greater than the continuum (this ratio is also observed for the high state). The presence of the  emission line may be associated with accretion, which continues in the low state but at a significantly lower rate. 
The excess of the H$\alpha$ emission  over the continuum is often observed for magnetic CVs in the low state \citep[e.g.][]{kennedy2020,kolbin2023}.
In this case, the continuum is dominated by the white dwarf emission.
Strong emission lines formed in the disk are also present in the spectra of 
WZ Sge-type dwarf novae in a quiescent state \citep[e.g.][]{amantayeva2021}.

We tried to estimate the accretion rate  $\dot{M}$ in the high and  low states of J1944  assuming that all the energy produced by the accretion is mainly emitted in X-rays, using the formula
\begin{equation}
    L_{X, {\rm bol}} \sim \frac{GM_1\dot{M}}{2R_1},
\end{equation}
where $L_{X, {\rm bol}}$ is the bolometric luminosity. Bolometric fluxes were obtained using the best fit to the X-ray spectra of J1944, $F^{h}_{X, {\rm bol}}\sim 7 \times10^{-12}$ \flux\ and $F^{l}_ {X, {\rm bol}}\sim 6\times10^{-13}$ \flux. As a result, the accretion rate in the low state is
\begin{equation}
\dot{M}^{l} 
\approx 3\times10^{-12} \frac{R_1}{10^4 \text{km}} \frac{{\rm M}_\odot}{M_1}\ \msun~\text{yr}^{-1},   
\end{equation}
while the accretion rate in the high state is an order of magnitude higher for any acceptable values from the mass $M_1$ and radius $R_1$ ranges of the white dwarf estimated above.


\subsection{Type of the cataclysmic variable}

As mentioned above, the presence of an accretion disk in J1944 excludes its classification as a (asynchronous) polar. Its properties are close to the VY Scl type CVs, which also demonstrate low and high states with typical flux variations in a range of 3 -- 5\m. The main differences are that VY Scl systems have typical orbital periods of 3 -- 6 hours \citep{VY-Scl2023}, white dwarf temperatures of more than 20,000 K, and the accretion rate in the high state by two orders of magnitude higher than for J1944. According to its characteristics, J1944 is most likely an intermediate polar.

The equivalent width ratio of HeII $\lambda$4686/H$\beta$ in the J1944 spectrum is less than the typical value $>$0.4 for magnetic CVs \citep{silber}. However, there are intermediate polars with  similar ratios, for example, DW Cnc\footnote{The source spectrum is available in the LAMOST survey \url{https://www.lamost.org/dr8/v2.0/spectrum/view?obsid=392716175}}. This is also a short-period system ($P_{orb}=86$ min) demonstrating  high and low states \citep{duffy2022}.


\subsection{Association with the $\gamma$-ray source \fgl}

J1944 is located in the error ellipse of the unidentified $\gamma$-ray  source \fgl, which raises the question of their possible association. $\gamma$-ray 
emission from CVs is extremely rarely detected. 
Only a few novae eruptions were 
observed in $\gamma$-rays \citep{gk-per,sokolovsky22,sokolovsky23}. Only for two rapidly rotating magnetic white dwarfs in binary systems, AE Aqr and AR Sco (the spin periods are 33 and 117 s, respectively), pulsations with the spin period of the white dwarf were possibly detected in the $\gamma$-rays \citep{fermi-pulsed-CV}. AE Aqr is an intermediate polar with a magnetic field of $B\sim10^6$ G, and AR Sco shows a number of properties that distinguish it from all known classes of CVs. The accretion rate in both systems is very low, an accretion disk is not formed: in AE Aqr -- due to the propeller effect, in AR Sco -- due to the strong magnetic field of the white dwarf ($B\sim10^8$ G).

Classifying J1944 as a CV makes the association with \fgl\ very unlikely, since there are no signs of a nova, there is an accretion disk and there are no such peculiar properties as for AE Aqr.

Nevertheless, we note that 4FGL J1943.9+2841 still could be a pulsar. Further multi-wavelength studies are needed to confirm 
this hypothesis.


\subsection{Conclusion}

Our results show that J1944 has a number of characteristic features which allow us  to classify it as an intermediate polar. It is remarkable that  in this case it joins 
a small group of intermediate polars with the shortest orbital periods, lying below the period gap \citep{11IPs-lstate}, and exhibiting transitions between the high and low brightness  states in the optical and X-rays. 
Currently, the low state has been detected in only eleven intermediate polars \citep[e.g.,][]{6-IPs-lowstate,11IPs-lstate}, and only two objects, FO Aqr and Swift J0746.3$-$1608 \citep{IP-accr-rate}, have been observed in X-rays in both the low and high states. Thus, the relatively bright source J1944 is a unique object for studying the evolution of similar systems simultaneously in the optical-UV and X-rays. Additional X-ray observations may reveal detailed information about the 8-min periodicity of the white dwarf emission. Optical-near-infrared spectroscopy of  J1944 in the low state will allow one to significantly clarify the parameters of the white dwarf and, possibly, the donor star, as well as the parameters of the binary system.


\subsection{Acknowledgments}

{\footnotesize This study used observations from the eROSITA telescope on board the SRG Observatory. The SRG observatory was manufactured by Roscosmos in the interests of the Russian Academy of Sciences represented by the Space Research Institute (IKI) within the framework of the Russian Federal Scientific Program with the participation of the German Aerospace Center (DLR). The X-ray telescope SRG/eROSITA was manufactured by a consortium of German institutes led by the Max Planck Institute for Extraterrestrial Astrophysics (MPE) with support from DLR. The SRG spacecraft was designed, manufactured, launched and managed by NPO Lavochkin and its subcontractors. Scientific data is received by a complex of deep space communication antennas in Bear Lakes, Ussuriysk and Baikonur and is financed by Roscosmos. The eROSITA telescope data used in this work were processed using eSASS software developed by the German eROSITA consortium and software developed by the Russian SRG/eROSITA telescope consortium. 

This work made use of data supplied by the UK Swift Science Data Centre at the University of Leicester.

This work has made use of data from the European Space Agency (ESA) mission
{\it Gaia} (\url{https://www.cosmos.esa.int/gaia}), processed by the {\it Gaia}
Data Processing and Analysis Consortium (DPAC,
\url{https://www.cosmos.esa.int/web/gaia/dpac/consortium}).
Funding for the DPAC
has been provided by national institutions, in particular the institutions
participating in the {\it Gaia} Multilateral Agreement.
Based on observations obtained with the Samuel Oschin Telescope 48-inch and the 60-inch Telescope at the Palomar Observatory as part of the Zwicky Transient Facility project. ZTF is supported by the National Science Foundation under Grants No. AST-1440341 and AST-2034437 and a collaboration including current partners Caltech, IPAC, the Weizmann Institute for
Science, the Oskar Klein Center at Stockholm University, the University of Maryland, Deutsches Elektronen-Synchrotron and Humboldt University, the TANGO Consortium of Taiwan, the University of Wisconsin at Milwaukee, Trinity College Dublin, Lawrence Livermore National Laboratories, IN2P3, University of Warwick, Ruhr University Bochum, Northwestern University and former partners the University of Washington, Los Alamos National Laboratories, and Lawrence Berkeley National Laboratories. Operations are conducted by COO, IPAC, and UW.
The Pan-STARRS1 Surveys (PS1) and the PS1 public science archive have been made possible 
through contributions by the Institute for Astronomy, the University of Hawaii, 
the Pan-STARRS Project Office, the Max-Planck Society and its participating institutes, 
the Max Planck Institute for Astronomy, Heidelberg and the Max Planck Institute 
for Extraterrestrial Physics, Garching, The Johns Hopkins University, Durham University, 
the University of Edinburgh, the Queen's University Belfast, 
the Harvard-Smithsonian Center for Astrophysics, 
the Las Cumbres Observatory Global Telescope Network Incorporated, 
the National Central University of Taiwan, the Space Telescope Science Institute, 
the National Aeronautics and Space Administration under Grant No. NNX08AR22G 
issued through the Planetary Science Division of the NASA Science Mission Directorate, 
the National Science Foundation Grant No. AST-1238877, the University of Maryland, 
Eotvos Lorand University (ELTE), the Los Alamos National Laboratory, 
and the Gordon and Betty Moore Foundation.
%

This paper makes use of data obtained as part of the INT Photometric H$\alpha$ Survey of the Northern Galactic Plane (IPHAS, www.iphas.org) carried out at the Isaac Newton Telescope (INT). The INT is operated on the island of La Palma by the Isaac Newton Group in the Spanish Observatorio del Roque de los Muchachos of the Instituto de Astrofisica de Canarias. All IPHAS data are processed by the Cambridge Astronomical Survey Unit, at the Institute of Astronomy in Cambridge. The bandmerged DR2 catalogue was assembled at the Centre for Astrophysics Research, University of Hertfordshire, supported by STFC grant ST/J001333/1.

The authors are grateful to TUBITAK, IKI, KFU and the Tatarstan Academy of Sciences for partial support in the use of RTT-150 (Russian-Turkish 1.5-m telescope in Antalya). Observations at telescopes of the Special Astrophysical Observatory of the Russian Academy of Sciences are carried out with the support of the Ministry of Science and Higher Education of the Russian Federation. The renovation of telescope equipment is currently
provided within the national project «Science and universities». Based upon observations carried out at the Observatorio Astronómico Nacional on the Sierra San Pedro Mártir (OAN-SPM), Baja California, México. We thank the daytime and night support staff at the OAN-SPM for facilitating and helping obtain our observations. The work was partially carried out within the framework of the state assignment of the Special Astrophysical Observatory of the Russian Academy of Sciences, approved by the Ministry of Science and Higher Education of the Russian Federation, as well as at the expense of the Strategic Academic Leadership Program of the Kazan (Volga Region) Federal University. \\

 The work of IFB, IMH, MAG, ENI, MVS, RIG, NAS was carried out with the help of a subsidy from the Ministry of Education and Science of the Russian Federation FZSM-2023-0015, allocated to Kazan Federal University to fulfill the state assignment in the field of scientific activity. \\

 The work of YuAS, AVK and DAZ (analysis of archival optical, ultraviolet and X-ray data, as well as analysis of data from the 2.1-m OAN-SPM telescope and the \eros\ telescope) was carried out within the framework of the state assignment of the Ioffe Institute number FFUG-2024-0002. DAZ thanks the Pirinem School of Theoretical Physics for hospitality. \\

AYK acknowledges the DGAPA-PAPIIT
grant IA105024 (planning optical observations with the OAN-SPM 2.1-m telescope and obtaining data).\\

The work of AIK (analysis of optical spectra, estimation of parameters of the binary system) was supported by the Russian Science Foundation grant No. 22-72-10064, https://rscf.ru/project/22-72-10064/. \\

}


\end{document}